\documentclass[twocolumn]{autart}    
\usepackage{amssymb}
\usepackage{graphicx}
\usepackage{mathrsfs}
\usepackage{epstopdf}
\usepackage{algorithm}
\usepackage{algorithmicx}
\usepackage{algpseudocode}
\usepackage{pifont}
\usepackage{array}
\usepackage{multirow}
\usepackage{color}
\usepackage{makecell}
\usepackage{amsmath}
\usepackage{amssymb}
\usepackage{mathrsfs}
\usepackage{pbox}
\usepackage{commath}
\usepackage{booktabs,caption}
\usepackage{soul}
\newcommand{\argmax}{\operatornamewithlimits{argmax}}

\makeatletter
\DeclareRobustCommand*\cal{\@fontswitch\relax\mathcal}
\makeatother
\newcommand{\bc}{\begin{center}}
\newcommand{\ec}{\end{center}}
\newcommand{\bi}{\begin{itemize}}
\newcommand{\ei}{\end{itemize}}
\newcommand{\ba}{\begin{array}}
\newcommand{\ea}{\end{array}}
\newcommand{\bal}{\begin{aligned}}
\newcommand{\eal}{\end{aligned}}
\newcommand{\beq}{\begin{equation}}
\newcommand{\eeq}{\end{equation}}
\newcommand{\bcs}{\begin{cases}}
\newcommand{\ecs}{\end{cases}}
\newcommand{\beqa}{\begin{eqnarray}}
\newcommand{\eeqa}{\end{eqnarray}}

\def\thetak{\theta^{(n)}}
\def\thetakp{\theta^{(n+1)}}
\def\Xk{{\bf X}_k}

\def\Xs{{\bf X}_s}

\def\Ys{{\bf Y}_s}
\def\Xsm{{\bf X}_{s-1}}

\def\Yk{{\bf Y}_k}
\def\Yr{{\bf Y}_r}
\def\Yok{{\bf Y}_{1:k}}
\def\Yokm{{\bf Y}_{1:k-1}}

\def\Yor{{\bf Y}_{1:r}}

\def\YT{{\bf Y}_T}
\def\XoT{{\bf X}_{0:T}}
\def\YoT{{\bf Y}_{1:T}}
\def\f{{\bf{f}}}

\def\v{{\bf{v}}}

\def\Xkm{{\bf X}_{k-1}}

\def\h{{\bf{h}}}
\def\nk{{\bf n}_k}

\def\vk{{\bf v}_k}

\def\hXk{{\bf \hat{X}}_{k|k}}
\def\hXkr{{\bf \hat{X}}_{k|r}}
\def\hXkT{{\bf \hat{X}}_{k|T}}

\def\hXsT{{\bf \hat{X}}_{s|T}}

\def\X{{\bf X}}
\def\Y{{\bf Y}}
\def\Yon{{\bf Y}_1}

\def\msec{{\rm MSE}}
\def\x{{\bf x}}
\def\y{{\bf y}}

\def\Pik{{\bf \Pi}_{k|k}}
\def\Pikr{{\bf \Pi}_{k|r}}

\def\Pikm{{\bf \Pi}_{k|k-1}}

\def\Pism{{\bf \Pi}_{s|s-1}}
\def\PikT{{\bf \Pi}_{k|T}}

\def\Pis{{\bf \Pi}_{s|s}}

\def\Pikmm{{\bf \Pi}_{k-1|k-1}}
\def\Pismm{{\bf \Pi}_{s-1|s-1}}

\def\Diks{{\bf \Delta}_{k|s}}

\def\Dis{{\bf \Delta}_{s|s}}
\def\Dism{{\bf \Delta}_{s|s-1}}
\def\Dismm{{\bf \Delta}_{s-1|s-1}}
\def\Dikm{{\bf \Delta}_{k|k-1}}

\def\Girp1{{\bf \Gamma}_{r+1|r}}

\def\DiT{{\bf \Delta}_{T|T}}

\def\Piz{{\bf \Pi}_{0|0}}

\def\Xz{{\bf X}_0}

\def\Mk{{M_k}}

\def\Ms{{M_s}}

\def\Tk{{T_k}}

\def\btk{{\mbox{\boldmath $\beta$}_k}}
\def\hbtk{{\hat{\mbox{\boldmath $\beta$}}}_k}
\def\bts{{\mbox{\boldmath $\beta$}_s}}

\def\Xk{{\bf X}_k}

\def\Yk{{\bf Y}_k}
\def\f{{\bf{f}}}
\def\v{{\bf{v}}}

\def\Xkm{{\bf X}_{k-1}}

\def\h{{\bf{h}}}
\def\nk{{\bf n}_k}

\def\vk{{\bf v}_k}

\def\X{{\bf X}}
\def\Y{{\bf Y}}
\def\Yon{{\bf Y}_1}

\def\msec{{\rm MSE}}
\def\x{{\bf x}}
\def\y{{\bf y}}

\def\Pik{{\bf \Pi}_{k|k}}
\def\Pikm{{\bf \Pi}_{k|k-1}}
\def\Pikmm{{\bf \Pi}_{k-1|k-1}}

\def\Piz{{\bf \Pi}_{0|0}}
\def\Xz{{\bf X}_0}

\def\Mk{{M_k}}
\def\Tk{{T_k}}
\def\Ts{{T_s}}

\def\btk{{\mbox{\boldmath $\beta$}_k}}
\def\bt{{\mbox{\boldmath $\beta$}}}

\def\tx{\tilde{\bf{x}}}

\def\tPi0{\tilde{\bf{\Pi}}_0}

\def\z{{\bf{z}}}

\def\hXkMS{\hXk^{\rm MS}}
\def\hXkrMS{\hXkr^{\rm MS}}
\def\hXkTMS{\hXkT^{\rm MS}}
\def\hXsTMS{\hXsT^{\rm MS}}

\def\hXkrMAP{\hXkr^{\rm MAP}}
\def\cXkr{{\cal X}_{k|r}}

\begin{document}
\begin{frontmatter}

\title{Particle Filters for
\\Partially-Observed Boolean Dynamical Systems} 


\author[Paestum]{Mahdi Imani}\ead{m.imani88@tamu.edu},    
\author[Paestum]{Ulisses Braga-Neto}\ead{ulisses@ece.tamu.edu}               

\address[Paestum]{Department of Electrical and Computer Engineering,Texas A\&M University,
College Station, TX, USA}  


\begin{abstract}                           Partially-observed
  Boolean dynamical systems (POBDS) are a general class of nonlinear models
  with application in estimation and control of Boolean processes 
  based on noisy and incomplete measurements. 
  The optimal minimum mean square
  error (MMSE) algorithms for POBDS state estimation,
  namely, the Boolean Kalman filter (BKF) and Boolean Kalman smoother
  (BKS), are intractable in the case of large systems, due to
  computational and memory requirements. To address this, we propose
  approximate MMSE filtering and smoothing algorithms based
  on the auxiliary particle filter (APF) method from sequential
  Monte-Carlo theory. These algorithms are used jointly with
  maximum-likelihood (ML) methods for simultaneous state and parameter
  estimation in POBDS models. In the presence of continuous
  parameters, ML estimation is performed using the
  expectation-maximization (EM) algorithm; we develop for this purpose
  a special smoother which reduces the computational complexity of the
  EM algorithm. The resulting particle-based adaptive filter is
  applied to a POBDS model of Boolean gene regulatory networks
  observed through noisy RNA-Seq time
  series data, and performance is assessed through a series of
  numerical experiments using the well-known cell cycle gene
  regulatory model.
\end{abstract}

\end{frontmatter}

\section{Introduction}
\label{sec:Intro}

Partially-observed Boolean dynamical systems consist of
a Boolean state process, also known as a Boolean network, observed
through an arbitrary noisy mapping to a measurement space
\cite{Brag:11,Brag:12,Brag:13,ImanBrag:15a,ImanBrag:ACC,BahaBrag:15,ImanBrag:16a,ImanBrag:17a,LeviBrag:17,ImanBrag:TCON,ImanBrag:TSP2}.
Instances of POBDSs abound in fields such as
genomics~\cite{Kauf:69}, robotics~\cite{Rolietal:11},
digital communication systems~\cite{Mess:90},
and more. The optimal recursive minimum mean-square error (MMSE) state
estimators for this model are called the Boolean Kalman Filter (BKF)
\cite{Brag:11} and the Boolean Kalman Smoother (BKS)
\cite{ImanBrag:15b}. These filters have many desirable properties; in
particular, it can be shown that the MMSE estimate of the state vector
provides both the MMSE and the maximum-a-posteriori (MAP) estimates of
each state vector component. Notice that the software tool
``BoolFilter"~\cite{BoolFilter} is available under R library for estimation and identification of
partially-observed Boolean dynamical systems.

However, for large systems with large number of state variables, the
computation of both the BKF and BKS becomes impractical due to large
computational and memory requirements. In \cite{Brag:13}, an
approximate sequential Monte-Carlo (SMC) algorithm was proposed to
compute the BKF using \textit{sequential importance resampling}
(SIR). By contrast, we develop here SMC algorithms for both the BKF
and fixed-interval BKS based on the more efficient {\em auxiliary
  particle filter} (APF) algorithm \cite{pitt1999filtering}.

The BKF and BKS require for their application that all system
parameters be known. In the case where noise intensities, the network
topology, or observational parameters are not known or only partially
known, an adaptive scheme to simultaneously estimate the state and
parameters of the system is required. An exact adaptive filtering
framework to accomplish that task was proposed recently
in~\cite{ImanBrag:16b}, which is based on the BKF and BKS in
conjunction with maximum-likelihood estimation of the parameters. In
this paper, we develop an accurate and efficient particle filtering
implementation of the adaptive filtering framework in
\cite{ImanBrag:16b}, which is suitable for large systems.

In the case where the parameter space is discrete (finite), the
adaptive filter corresponds to a bank of particle filters in parallel,
which is reminiscent of the multiple model adaptive estimation (MMAE)
procedure for linear systems \cite{maybeck1995performance}. If the
parameter space is continuous, then a particle-based version of the
Expectation Maximization (EM) algorithm \cite{DempLairRubi:77} is
developed. The computational complexity of EM method arises from three
main parts:
\begin{enumerate}
\item The computational complexity of applying smoothing at the E-step.
\item Memory necessary to store the required matrices and vectors
  (e.g. the posterior probability vectors) from the E-Step to the M-Step.
\item The complexity of each iteration in the M-step in which several function evaluations are required.
\end{enumerate}
Our proposed particle-based implementation addresses each of the above
issues.\nocite{nozari2015differentially}

Our application of interest in this paper is to model Boolean gene
regulatory networks~\cite{Kauf:69,Shmuetal:02b} observed through a
single time series of RNA-seq data~\cite{MargBahl:10}. Using the POBDS
model, we employ the proposed approximate adaptive ML algorithm to
estimate the gene expression state simultaneously to the inference of
the network topology and noise and expression parameters. Performance
is assessed through a series of numerical experiments using the
well-known cell-cycle gene regulatory model \cite{Fauretal:06}. The
influence of transition noise, expression parameters, and RNA-seq
measurement noise (data dispersion) on performance is studied, and the
consistency of the adaptive ML filter (i.e., convergence to true
parameter values) is empirically established.

The article is organized as follows. In Section~\ref{sec:BDS}, the
POBDS signal model and the Boolean Kalman Filter and Boolean Kalman
Smoother are reviewed, while in Section~\ref{sec:PF}, a detailed
description of the APF-based filtering and smoothing algorithms
proposed in this paper is provided. In Section~\ref{sec:PE}, the
particle-based ML adaptive filter is developed for discrete and
continuous parameter spaces. A POBDS model for gene regulatory
networks observed though RNA-seq measurements is reviewed in
Section~\ref{sec:GRN}. Results for the numerical experiments with the
cell-cycle network are presented in Section~\ref{sec:NE}. Finally,
Section~\ref{sec:conc} contains concluding remarks.

\section{Optimal State Estimators for POBDS}
\label{sec:BDS}

In this section, we review the POBDS model and exact
algorithms for computation of its optimal state estimators. For more
details see
\cite{Brag:11,ImanBrag:15a,ImanBrag:ACC,ImanBrag:15b}. For
a proof of optimality of the BKF, see \cite{ImanBrag:16b}.


We assume that the system is described by a {\em 
state process} $\{\Xk; k=0,1,\ldots\}$, where $\Xk \in \{0,1\}^d$ is a Boolean vector of size
$d$. The sequence of states is observed indirectly through the {\em observation
  process} $\{\Yk ; k=1,2,\ldots\}$, where $\Yk$ is a vector of
(typically non-Boolean) measurements. The states are assumed to be
updated and observed at each discrete time through the following nonlinear signal model:
\beq
\bal
  \Xk &\,=\, \f_k\left(\Xkm\right) \:\:\oplus\:\: \nk\quad \textrm{(state model)} \\[1ex]
  \Yk &\,=\, \h_k\left(\Xk,\vk\right) \quad \textrm{(observation model)}
\label{eq-sgnmodel}
\eal
\eeq
for $k=1,2,\ldots$ Here, $\nk \in \{0,1\}^d$ is Boolean transition
noise, ``$\oplus$'' indicates componentwise modulo-2 addition,
$\f_k: \{0,1\}^{d} \rightarrow \{0,1\}^d$ is a Boolean function, called
the {\em network function}, whereas $\h_k$ is a general function mapping
the current state and observation noise $\vk$ into the measurement
space, for $k=1,2,\ldots$. The noise processes $\{\nk, \vk; k=1,2,\ldots\}$ are assumed to
be ``white'' in the sense that the noises at distinct time points are
independent random variables.  It is also assumed that the noise
processes are independent from each other and from the initial state
$\Xz$; their distribution is otherwise arbitrary. 

We will assume further that the Boolean process noise $\nk$ is {\bf
  zero-mode}, i.e., $\nk = {\bf 0}$ is the most probable value of the
noise vector at each time $k$. This implies that the most likely value
of $\Xk$ at each time $k$ is $\f_k(\Xkm)$ --- this could be seen as
the counterpart of the zero-mean noise assumption in continuous
state-space models. As is the case with nonzero mean noise, nonzero
mode noise introduces a systematic error component, which can always
be removed by moving it into the function $\f_k$. Hence, the state
model in (\ref{eq-sgnmodel}) is a general model for a first-order
Markov Boolean stochastic process. For a specific example, which will
be adopted in Section~\ref{sec:NE}, one has $P(\nk(i)=1)= p$, for
$i= 1,\ldots,d$, and $k=1,2,\ldots$ independently for $i \neq j$. In
this case, $\nk$ is zero-mode if and only if $p \leq 1/2$. The
systematic bias introduced in the case $p > 1/2$ can be removed by
considering the equivalent state process with
$\f_k^\prime = {\bf{1}} - \f_k$, where ${\bf{1}}$ is the vector with
all components equal to 1, and $p^\prime = 1 - p$, i.e., by moving the
systematic bias into the model. Therefore, one effectively needs only
to consider the case~$p \leq 1/2$.


A  Boolean estimator $\hXkr$ predicts the Boolean state $\Xk$
based on the sequence of observations
$\Yor = (\Yon,\ldots,\Yr)$. 
The estimator $\hXkr$ is called a {\em filter},
{\em smoother}, or {\em predictor} according to whether
$k = r$, $k < r$, or $k >  r$, respectively.  
The set of all Boolean estimators for a given $k$ and $r$ shall be denoted by
$\cXkr$. The (conditional) mean-square error (MSE) of $\hXkr$ given $\Yor$ is:
\beq
  \msec(\hXkr \mid \Yor)\,=\, E\left[||\hXkr - \Xk||^2 \mid  \Yor\right]\,.
\label{eq-msec}
\eeq
We would like to obtain the {\em Boolean MMSE estimator}, i.e., a
Boolean estimator $\hXkrMS$ such that
\beq
  \hXkrMS \,=\, \arg\!\! \min_{\hXkr \in \cXkr} \!\! \msec(\hXkr \mid \Yor)\,,
\label{eq-argmin}
\eeq
at each value of $\Yor$ (so that it also minimizes the frequentist
expected MMSE over all possible realizations of $\Yor$).  For a Boolean vector $\v \in
\{0,1\}^d$, define the {\em binarized vector} $\overline{\v}$, such
that $\overline{\v}(i) = 1$ if $\v(i) > 1/2$ and $\overline{\v}(i) = 0$ otherwise, for
$i=1,\ldots,d$, the {\em complement vector}
$\v^c$, such that $\v^c(i) = 1-\v(i)$, for $i=1,\ldots,d$, and the {\em $L_1$-norm} $||\v||_1 =
\sum_{i=1}^{d} |\v(i)|$. It can be
proved \cite{ImanBrag:16b} that the solution to (\ref{eq-argmin}) is given by 
\beq
  \hXkrMS \,=\, \overline{E\left[\Xk \mid \Yor\right]}\,,
\label{eq:BKF6}
\eeq
with optimal MMSE
\beq
\bal
  & \msec(\hXkrMS\mid \Yor) \\
  & \:=\, \big|\big|\min\big\{E\left[\Xk \mid
      \Yok\right],E\left[\Xk \mid \Yok\right]^c\!\big\}\,\big|\big|_1,
\eal
\label{eq:BKF6a}
\eeq
where the minimum is computed componentwise.

The optimal Boolean MMSE estimator will be called a {\em Boolean
  Kalman Filter} (BKF), {\em Boolean Kalman Smoother} (BKS), or {\em
  Boolean Kalman Predictor} (BKP), according to whether $k = r$,
$k < r$, or $k > r$, respectively. The terminology is justified by the
fact that we seek the MMSE estimator for a nonstationary process, as in the case of the
classical Kalman Filter, as opposed to, say, the Maximum-A-Posteriori
(MAP) estimator, which is more common in discrete
settings. Interestingly, we can show that each component of the MMSE
estimator $\hXkrMS(i)$ is {\em both} the MMSE and the MAP estimator of
the corresponding state variable $\Xk(i)$, for $i=1,\ldots,d$. Perhaps
surprisingly, the MAP estimator $\hXkrMAP$ does not enjoy in general
the property that $\hXkrMAP(i)$ is the MAP estimator of $\Xk(i)$, for
$i=1,\ldots,d$. In cases where optimal estimation performance for each
component of $\hXkr$ is required (e.g., estimating the state of each
gene in a gene regulatory network), then this is an important
distinction.


Let $(\x^1,\ldots,\x^{2^d})$ be an arbitrary enumeration of the
possible state vectors, define the state conditional probability
distribution vector $\Pikr$ of length $2^d$ via
\beq
  \Pikr(i) \,=\, P\left(\Xk = \x^i \mid  \Yor\right)\,, \:\:i=1,\ldots,2^d,
\eeq 
and let $A \,=\, \left[\x^1 \cdots \x^{2^d}\right]$ be a matrix of
size $d \times 2^d$. Then it is clear
that $E\left[\Xk \mid \Yor\right] = A \Pikr$, so 
it follows from (\ref{eq:BKF6}) and (\ref{eq:BKF6a}) that
\beq
\hXkrMS \,=\, \overline{A \Pikr}\,,
\label{eq-estkr}
\eeq
with optimal MSE
\beq
  \msec(\hXkrMS\mid \Yor) \,=\, ||\min\{A\Pikr, (A\Pikr)^c\}||_1.
\label{eq-MSEkr}
\eeq

The distribution vector $\Pikr$ can be computed by a matrix-based procedure similar
to the ``forward-backward'' algorithm \cite{Rabi:89}. Briefly, let
$\Mk$ of size $2^d \times 2^d$ be the transition matrix of the Markov
chain defined by the state model:
\beq
\bal
   & (\Mk)_{ij} \,=\, P(\Xk = \x^i \mid \Xkm = \x^j)\\
   & \:\:=\, P\left(\nk \,=\, \f(\x^j) \oplus \x^i \right),\:\: i,j = 1,\ldots,2^d.
\eal
\eeq
Additionally, given a value of the
observation vector $\Yk$ at time $k$, let $\,\Tk(\Yk)$
be a diagonal matrix of size $2^d \times 2^d$ defined by:
\begin{equation}
  \left(T_k(\Yk)\right)_{ii} \,=\, p\left(\Yk \mid \Xk = \x^i\right),\:\: i = 1,\ldots,2^d\,,
\label{eq:BKF4a}
\end{equation}
where $p$ is either a probability density or a mass function, according to the
nature of the measurement $\Yk$. 

At this point, we distinguish two cases. The first is a recursive
implementation of the Boolean Kalman Filter (BKF), which does not need
a backward iteration, and can be iterated forward as new observations
arrive, for as long as desired. In this case, we use
(\ref{eq-estkr}) and (\ref{eq-MSEkr}) with $r = k$ to get the optimal filter
estimator and its minimum MSE~\cite{Brag:11}. The entire procedure is 
given in~Algorithm~\ref{alg:BKF}.

\begin{algorithm}
\caption{BKF: Boolean Kalman Filter}
\begin{algorithmic}[1]
  \small
\State Initialization: $(\Piz)_i \,=\, P\left(\Xz = \x^i\right)$, for $i=1,\ldots,2^d$.  
\vspace{01ex}
\noindent
For $k = 1,2,\ldots$, do:

\State Prediction: $\Pikm \,=\, \Mk\,\Pikmm$.
\vspace{1ex}

\State Update: $\btk \,=\, \Tk(\Yk)\, \Pikm$.
\vspace{1ex}

\State Normalization: $\Pik \,=\, \btk/||\btk||_1$. 

\vspace{1ex}
\State MMSE Estimator Computation: 
$\hXkMS \,=\, \overline{A \Pik}\,.$
\vspace{1ex}
\State 
$\msec(\hXkMS \mid \Yok) \,=\, ||\min\{A\Pik, (A\Pik)^c\}||_1.$
\end{algorithmic}\label{alg:BKF}
\end{algorithm}

The second case is a {\em fixed-interval} Boolean Kalman Smoother,
where a fixed batch of observations $\YoT = (\Yon,\ldots,\YT)$ of length
$T$ is available, and it is desired to obtain estimates of the state
at all points in the interval $k=1,\ldots,T$. 
In this case, a backward iteration will be needed (unless $k =
T$). Define the probability
distribution vector $\Diks$ of length $2^d$ via
\beq
  \Diks (i) \,=\, p\left(\Y_{s+1},\ldots,\YT \mid \Xk = \x^i \right)\,, \:\:i=1,\ldots,2^d,
\eeq 
for $s=0,\ldots,T$, where $\DiT$ is defined to be
${\bf 1}_{d \times 1}$, the vector with all components equal to 1.
It can be shown that
\beq
  \PikT \,=\, \frac{\Pikm \,\bullet\, \Dikm}{||\Pikm \,\bullet\,
    \Dikm||_1}\,,
\eeq
where ``$\,\bullet\,$'' denotes componentwise vector multiplication.
We then use (\ref{eq-estkr}) and (\ref{eq-MSEkr}) with $r = T$ to get the optimal smoothed
estimator and its minimum MSE~\cite{ImanBrag:15b}. The entire procedure is 
given in~Algorithm~\ref{alg:BKS}.

\begin{algorithm}
\caption{BKS: Fixed-Interval Boolean Kalman Smoother}
\begin{algorithmic}[1]
  \small


\State Initialization: 
$(\Piz)_i \,=\, P\left(\Xz = \x^i\right)$, for $i=1,\ldots,2^d$.  

\vspace{1ex}
\noindent
\underline{\em Forward Probabilities:} For $s = 1,\ldots,T$, do:

\vspace{1ex}
\State Prediction: $\Pism \,=\, \Ms\,\Pismm$.
\vspace{1ex}

\State Update: $\bts \,=\, \Ts(\Ys)\, \Pism$.
\vspace{1ex}

\State Normalization: $\Pis \,=\, \bts/||\bts||_1$. 

\vspace{1.5ex}
\noindent
\underline{\em Backward Probabilities:} For $s = T,T-1,\ldots,1$, do:

\vspace{1ex}
\State Update: $\Dism \,=\, \Ts(\Ys)\,\Dis$ \, (with $\DiT = {\bf{1}}$).

\vspace{1ex}
\State Prediction: $\Dismm \,=\, \Ms^T\,\Dism$.

\vspace{1.5ex}
\noindent
\underline{\em MMSE Estimator Computation:} 
For $k = 1,\ldots,T$, do:

\vspace{1ex}
\State $\PikT \,=\, (\Pikm \,\bullet\, \Dikm)/||\Pikm \,\bullet\,
    \Dikm||_1\,.$

\vspace{1ex}
\State$\hXkTMS \,=\, \overline{A \PikT}\,.$

\vspace{0.6ex}
\State $\msec(\hXkTMS \mid \YoT) \,=\, ||\min\{A\PikT, (A\PikT)^c\}||_1.$



\end{algorithmic}\label{alg:BKS}
\end{algorithm}


\section{Particle Filters for State Estimation}
\label{sec:PF}

When the number of states is large, the exact computation of the BKF
and the BKS becomes
intractable, due to the large size of the matrices involved, which
each contain $2^{2d}$ elements, and approximate methods must be
used, such as sequential Monte-Carlo methods, also
known as {\em particle filter algorithms}
which have been successfully applied in various fields~\cite{ghoreishi2016compositional,godsill2012monte,ghoreishi2016uncertainty,doucet2000sequential,friedman2017quantifying}.  In the next
subsections, we describe particle filter implementations of the
BKF and BKS.  

\subsection{Auxiliary Particle Filter Implementation of the BKF (APF-BKF)}
\label{sec:APF-BKF}

The basic algorithm to perform particle filtering is called
\textit{sequential importance resampling} (SIR). Importance sampling
is used when direct sampling of the target distribution is
difficult. The idea is to approximate the target distribution $p(\x)$
using sample points (``particles'') $\{\x_i\}_{i=1}^N$ drawn from a
proposal distribution $q(\x)$, which is easier to sample than the
target distribution. The discrepancy created by sampling from $q(\x)$
instead of $p(\x)$ is compensated by weighting each particle. 
After a few iterations of the algorithm, a condition in usually
 reached where only few of the particles have significant weights,
 whereas most particles have negligible weight. To address this
 degeneracy problem, SIR performs resampling of the particles, whereby
 a fresh set of particles is drawn (with replacement) from the
 approximate current posterior distribution.

 The original particle filtering implementation of the BKF
 in~\cite{Brag:13} was based on the SIR algorithm; we therefore call
 it the SIR-BKF algorithm. We present here a more sophisticated
 implementation based on the {\em Auxiliary Particle Filter} (APF) of
 \cite{pitt1999filtering}.  The APF algorithm can be seen as a
 variation of SIR, and is thus also known as auxiliary SIR (ASIR).
 Basically, APF is a look-ahead method that at time step $k-1$ tries
 to predict the location of particles with high probability at time
 $k$, with the purpose of making the subsequent resampling step more
 efficient. Without the look-ahead, the basic SIR algorithm blindly
 propagates all particles, even those in low probability regions. As
 put in \cite{WhitJoha:11}, ``it is natural to ask whether it is
 possible to employ knowledge about the next observation {\em before}
 resampling to ensure that particles which are likely to be compatible
 with that observation have a good chance of surviving.''

The APF algorithm augments the state vector to $(\Xk,\zeta_k)$, where
$\zeta_k$ is an auxiliary variable. Particles are drawn from the
filtering distribution $P(\X_k,\zeta_k \mid \Y_k)$ (to be specified
below), and the auxiliary variable is simply dropped to obtain
particles from $P(\X_k \mid \Y_k)$. 
Given particles $\{\x_{k-1,i}\}_{i=1}^N$ at time $k-1$, with associated weights
$\{W_{k-1,i}\}_{i=1}^N$, the APF algorithm defines
\beq
P(\Xk,\zeta_k \!\mid\! \Yk)\,\propto\,p(\Yk \!\mid\! \Xk)\,P(\Xk\!\mid\!
\x_{k-1,\zeta_k})\,W_{k-1,\zeta_k}\,,
\label{eq:apfd}
\eeq
for $\zeta_k = 1,\ldots,N$. The auxiliary variable functions thus as
an index for the particles at the previous time point. As will be seen
below, 
sampling from (\ref{eq:apfd}) will have the effect of ``selecting'' the
particles that are compatible with the observation at time $k$.

One can sample from (\ref{eq:apfd}) by using SIR on the following 
approximation:
\beq
P(\Xk,\zeta_k \!\mid\! \Yk)\,\propto\,p(\Yk \!\mid\! \mu_{k,\zeta_k})\,P(\Xk\!\mid\!
\x_{k-1,\zeta_k})\,W_{k-1,\zeta_k}\,,
\label{eq:apfd2}
\eeq
for $\zeta_k = 1,\ldots,N$, where $\mu_{k,i}$ is a characteristic of
$\Xk$ given $\x_{k-1,i}$, which can be the mean, the mode or even a sample
from $P(\Xk\mid \x_{k,i})$ \cite{pitt1999filtering}. In our
implementation, we use the mode:
\beq
\bal
\mu_{k,i}\,&=\,\text{Mode}[\Xk \mid \x_{k-1,i}]\\
\,&=\,\text{Mode}[\f(\x_{k-1,i}) \oplus \nk]\,=\,\f(\x_{k-1,i})\,,
\label{eq-mean}
\eal
\eeq
for $i=1,\ldots,N$, where we used~(\ref{eq-sgnmodel}). The
approximation on the right is accurate as long as the ``noise
intensity'' is low, i.e., the probability of nonzero $\nk$ is small.

Sampling from (\ref{eq:apfd}) is done in two steps. In the first step,
$\{\mu_{k,i}\}_{i=1}^N$ is obtained from the particles
$\{\x_{k-1,i}\}_{i=1}^N$ using (\ref{eq-mean}) and the {\em
  first-stage} weights $\{V_{k,i}\}_{i=1}^N$ are computed as:
\beq
V_{k,i} \,=\, p(\Yk\mid \mu_{k,i}) W_{k-1,i}\,,
\eeq
for $i=1,\ldots,N$. In the second step, the auxiliary variables
$\{\zeta_{k,i}\}_{i=1}^N$ (i.e., the indices of the selected
particles) are obtained as a sample from the discrete distribution
defined by $\{V_{k,i}\}_{i=1}^N$ (after proper normalization). For
example, if $N=4$ and $V_{k,1} = V_{k,2}$, $V_{k,3} = V_{k,4}$, and
$V_{k,1} = 2V_{k,3}$, then the indices
$\zeta_{k,0},\ldots,\zeta_{k,4}$ will be independent and each will be
twice as likely to be 1 or 2 than 3 or 4. We denote this by
$\{\zeta_{k,i}\}_{i=1}^N \sim {\rm Cat}(\{V_{k,i}\}_{i=1}^N)$,
where ``Cat'' stands for the categorical (discrete) distribution. 

Finally, the new particles $\{\x_{k,i}\}_{i=1}^N$ and associated {\em
  second-stage} weights $\{\tilde{W}_{k,i}\}_{i=1}^N$ can be obtained as follows:
\beq
\x_{k,i}\, =\, \mu_{k,\zeta_{k,i}}\oplus \nk \,\sim\, P(\X_k \mid \x_{k-1,\zeta_{k,i}})\,,
\eeq
\beq
\tilde{W}_{k,i}\,=\,\frac{p(\Y_k\mid\x_{k,i})}{p(\Y_k\mid\mu_{k,\zeta_{k,i}})}\,.
\eeq
It can be shown that the unbiased estimator of the unnormalized posterior probability at each time step can be obtained by~\cite{pitt2002smooth}
\beq
||\hbtk||_1\,=\,\left(\frac{1}{N}\,\sum_{i=1}^N V_{k,i}\right)\,\left(\frac{1}{N}\,\sum_{i=1}^N \tilde{W}_{k,i}\right)\,.
\eeq
This quantity will be needed in Section~\ref{sec:PE} when the particle
filter for maximum-likelihood adaptive estimation is discussed.

Given the normalized second-stage weights $W_{k,i}=\tilde{W}_{k,i}/\sum_{j=1}^N \tilde{W}_{k,j},\,i=1,\ldots,N$,
one can write
\beq
  E[\Xk \mid \Y_{1:k}] \,\approx\, \z_k = \sum_{i=1}^N W_{k,i}\,\x_{k,i}\,.
\eeq
From (\ref{eq:BKF6}) and (\ref{eq:BKF6a}), it follows that the MMSE state estimate and conditional MSE at time step $k$ are approximated as: 
\beq
\hXkMS\,=\, \overline{\z_k}\,,
\eeq
with optimal MMSE
\beq
\msec(\hXkMS,\Y_{1:k}) \,=\, ||\min\{\z_k, \z_k^c\}||_1.
\eeq
The entire procedure of APF-BKF is summarized in Algorithm~\ref{alg:APF}.


\begin{algorithm}
\caption{APF-BKF: Auxiliary Particle Filter implementation of the
  Boolean Kalman Filter}
\label{alg:APF}
\begin{algorithmic}[1]
\State $\x_{0,i} \sim {\bf{\Pi}}_{0|0}, W_{0,i}=1/N$, for $i=1,\ldots,N$.\vspace{0.8ex}
\For{$k =1,2,\ldots,$} \vspace{0.8ex}
 \For{$i =1$ to $N$}\vspace{0.8ex}
\State $\mu_{k,i}\,=\,\f(\x_{k-1,i})$.\vspace{0.8ex}
\State ${V}_{k,i}\,=\,p(\Y_k\mid\mu_{k,i})\,W_{k-1,i}$.\vspace{0.8ex}
\EndFor\vspace{0.8ex}
\State $\{\zeta_{k,i}\}_{i=1}^N \sim {\rm Cat}(\{V_{k,i}\}_{i=1}^N)$.\vspace{.8ex}
\For{$i =1$ to $N$}\vspace{0.8ex}
\State $\x_{k,i}\,=\,\mu_{k,\zeta_{k,i}}\oplus \nk$.\vspace{.8ex}
\State $\tilde{W}_{k,i}\,=\,\frac{p(\Yk\mid\x_{k,i})}{p(\Yk\mid\mu_{k,\zeta_{k,i}})}$.\vspace{0.8ex}
\EndFor\vspace{.8ex}
 \State $\Vert \hbtk \Vert_1\,=\,\left(\frac{1}{N} \sum_{i=1}^N V_{k,i}\right)\,\left(\frac{1}{N} \sum_{i=1}^N \tilde{W}_{k,i}\right)$.\vspace{.8ex}
\State $W_{k,i}=\tilde{W}_{k,i}/\sum_{j=1}^N \tilde{W}_{k,j},\,i=1,\ldots,N$.\vspace{0.8ex}
\State $\z_k = \sum_{i=1}^N W_{k,i}\,\x_{k,i}$.\vspace{.8ex}
\State $\hXkMS \,=\, \overline{\z_k}$.\vspace{.8ex}
\State $\msec(\hXkMS,\Y_{1:k}) \,=\, ||\min\{\z_k, \z_k^c\}||_1$.\vspace{.8ex}
\EndFor
\end{algorithmic}
\end{algorithm}

\subsection{Auxiliary Particle Filter Implementation of the BKS (APF-BKS)}

There are a few different approximate Monte-Carlo smoothing methods in
the literature of nonlinear and non-Gaussian
systems~\cite{kitagawa1996monte,doucet2000sequential,barembruch2009approximate}.
It should be noted that some of these particle smoother methods suffer
from degeneracy problems or can only be applied in a few special
conditions (such as MC with good forgetting properties). We follow an
approach similar to the well-known fixed-interval smoother
of~\cite{hurzeler1998monte} to approximate the Boolean Kalman
Smoother.

As described in Section~\ref{sec:BDS}, a fixed-interval smoother is a
forward-backward method, such that the filtering distributions $\Pik$
for $k=0,1,\ldots,T$ are computed in the forward step, and the smoothed
distributions $\PikT$ are found in a backward step.  The forward
process is obtained here by running the APF-BKF algorithm of
Section~\ref{sec:APF-BKF}, while the backward process is performed by
correcting the filtering weights in the backward iteration. We explain
next how the backward step is applied efficiently.

First, assume $\{\x_{k,i},W_{k,i}\}$, $k=0,\ldots,T$ are the forward
particles and weights obtained by the APF-BKF algorithm for the
sequence of measurements $\Y_{1:T}$.  Due to the finite number of
states in the POBDS, one can compute unique particles and their
associated weights at different time steps as:
\beq\label{eq-uniqF}
\{\x^u_{k,i},W^u_{k,i}\}_{i=1}^{F_k} \xleftarrow{\text{Unique}} \{\x_{k,i},W_{k,i}\}_{i=1}^{N}\,,\,k=0,\ldots,T.
\eeq 
where ${F_k}$ is the number of unique forward particles, and
$\x^u_{k,i}$ is the $i$th unique particle with aggregated weight
$W^u_{k,i}$, both at time step $k$.

The backward process is based on the following equation: 
\beq\label{eq:Smoo1}
\bal
P(\X_s,&\,\X_{s+1}\mid\Y_{1:T})\,\\
=&\,P(\X_s\mid\X_{s+1},\Y_{1:T})\,P(\X_{s+1}\mid\Y_{1:T})\\
=&\,P(\X_s\mid\X_{s+1},\Y_{1:s})\,P(\X_{s+1}\mid\Y_{1:T})\\
=&\frac{P(\X_{s+1}\mid\X_s) P(\X_s\mid\Y_{1:s})P(\X_{s+1}\mid\Y_{1:T})}{P(\X_{s+1}\mid\Y_{1:s})}\,,
\eal
\eeq
where $s<T$ and $P(\X_{s+1}\mid\Y_{1:T})$ is the smoothed distribution
at time step $s+1$. The summation over $\X_{s+1}$ in both sides of
equation~(\ref{eq:Smoo1}) results in 
\beq\label{eq:reweight1}
\bal
P(\X_{s}\mid&\Y_{1:T})\,=\,P(\X_s\mid\Y_{1:s})\,\\
&\,\times\sum_{\X_{s+1}} \frac{P(\X_{s+1}\mid\X_s)\,P(\X_{s+1}\mid\Y_{1:T})}{P(\X_{s+1}\mid\Y_{1:s})}\,.
\eal
\eeq
As we mentioned before, the filter and smoother estimate at final time
$T$ are the same. Therefore, the smoothed weights $W_{T|T,i}$ are
defined in the same way as the forward unique weights $W_{T,i}^u$, so that
\beq
P(\X_{T}\mid\Y_{1:T})\approx \sum_{i=1}^{F_{T}} W_{T|T,i}\, \delta_{\x^u_{T,i}}\,.
\eeq
Now, using equation~(\ref{eq:reweight1}), the the smoothed weights 
at time $s<T$ can be obtained as:
\beq\label{eq:Smoo2}
\bal
W_{s|T,j}\,=\,W^u_{s,j}\, \sum_{i=1}^{F_{s+1}} \,\frac{P(\x^u_{s+1,i}\mid\x^u_{s,j})\,W_{s+1|T, i}}{\sum_{l}^{F_{s}}\,P(\x^u_{s+1,i}\mid\x^u_{s,l})\, W^u_{s,l}}\,.
\eal
\eeq
The smoothed weights are obtained by solving equation~(\ref{eq:Smoo2})
in a backward fashion using the terminal condition
$W_{T|T,j}=W_{T,j}^u$, $j=1,\ldots,F_T$.  The computational complexity of
equation~(\ref{eq:Smoo2}) is of order $O(F_s\times F_{s+1})$ which can
be much smaller than $O(N\times N)$ in practice.

Using the smoothed weights, we can write
\beq
  E[\Xs \mid \Y_{1:T}] \,\approx\, \z_s\, = \,\sum_{i=1}^{F_s} W_{s|T,j}\, \x_{s,i}^u\,.
\eeq
From (\ref{eq:BKF6}) and (\ref{eq:BKF6a}), it follows that the MMSE state estimate and conditional MSE at time step $k$ are approximated as: 
\beq
\hXsTMS\,=\, \overline{\z_s}\,,
\eeq
with optimal MMSE
\beq
\msec(\hXsTMS,\Y_{1:T}) \,=\, ||\min\{\z_s, \z_s^c\}||_1.
\eeq
The entire procedure is summarized in Algorithm~\ref{alg:reweight}.

\begin{algorithm}
\caption{APF-BKS: Auxiliary Particle Filter implementation of the fixed-interval
  Boolean Kalman Smoother}
\label{alg:reweight}
\begin{algorithmic}[1]
\small
\State Run the APF-BKF for the sequence of measurements $\Y_{1:T}$ to obtain $\{\x_{k,i},W_{k,i}\}_{i=1}^N$, $k=0,\ldots,T$. \vspace{0.8ex}
\State $\{\x^u_{k,i},W^u_{k,i}\}_{i=1}^{F_k} \xleftarrow{\text{Unique}} \{\x_{k,i},W_{k,i}\}_{i=1}^{N},k=0:T$.\vspace{0.8ex}
\State Set $W_{T|T,i}\,=\,W^u_{T,i}$, for $i=1,\ldots,F_T$.\vspace{0.8ex}
\For{$s =T-1$ to $0$} \vspace{0.8ex}
\For{$j =1$ to $F_s$} \vspace{0.8ex}
\State\vspace{-5ex} \beq\nonumber
\bal
W_{s|T,j}\,=\,W^u_{s,j}\, \sum_{i=1}^{F_{s+1}}\,\frac{P(\x^u_{s+1,i}\mid\x^u_{s,j})\, W_{s+1|T, i}}{\sum_{l}^{F_s} \,P(\x^u_{s+1,i}\mid\x^u_{s,l})\,W^u_{s,l}}\,
\eal
\eeq\vspace{-3ex}
\EndFor\vspace{0.8ex}
\State $ \z_s\,=\,\sum_{i=1}^{F_s} W_{s|T,i}\,\x^u_{s,i}\,$.\vspace{.8ex}
\State $ \hXsTMS \,=\, \overline{\z_s}$.\vspace{.8ex}
\State $\msec(\hXsTMS \mid \YoT) \,=\, ||\min\{\z_s, \z_s^c\}||_1.$\vspace{.8ex}
\EndFor
\end{algorithmic}
\end{algorithm}

This particle smoother is an efficient method for state estimation, as
will be shown in Section~\ref{sec:NE}, but it is not appropriate for
parameter estimation, as we will argue in the next section. A
different particle smoother will be used in the next section to
perform continuous parameter estimation.

\section{Particle Filters For Maximum-Likelihood Adaptive Estimation}
\label{sec:PE}
 
Suppose that the nonlinear signal model in (\ref{eq-sgnmodel}) is
incompletely specified. For example, the deterministic functions
$\f_k$ and $\h_k$ may be only partially known, or the statistics of
the noise processes $\nk$ and $\vk$ may need to be estimated.  By
assuming that the missing information can be coded into a
finite-dimensional vector parameter $\theta \in \Theta$, where
$\Theta$ is the parameter space, we propose next particle filtering
approaches for simultaneous state and parameter estimation for
POBDS. For simplicity and conciseness, we consider two cases: a
Boolean Kalman Filter algorithm with finite (i.e., discrete) $\Theta$
and a Boolean Kalman Smoother algorithm with $\Theta \subseteq R^m$,
but the algorithms can be modified and even combined to perform other
estimation tasks. Exact algorithms for such filters can be found in
\cite{ImanBrag:16b}.

\subsection{APF Implementation of the Discrete-Parameter ML Adaptive BKF (APF-DPMLA-BKF)}
\label{subsec:MMAE}

In this case, $\Theta =\{\theta_1,\theta_2,\ldots,\theta_M\}$. 
Given the observations
$\Yok = \{\Y_1,\ldots,\Y_k\}$ up to time $k$, 
the log-likelihood function can be written as 
\beq
\bal
L_k(\theta_i)& \,=\, \log\, p_{\theta_i}(\Yok) \\
&\,=\, \log\, p_{\theta_i}(\Yk \mid \Yokm)\,+\, \log\, p_{\theta_i}(\Yokm) \\
&\,=\, \log\, p_{\theta_i}(\Yk \mid \Yokm) \,+\, L_{k-1}(\theta_i)\,,
\eal
\label{eq-Lrec}
\eeq
for $i=1,\ldots,M$, where
$\Vert\bt_k^{\theta_i}\Vert_1 = p_{\theta_i}(\Yk \mid \Yokm)$ can be
approximated by running the APF-BKF algorithms
discussed in the previous section tuned to parameter vector $\theta_i$.  

The approximate log-likelihood is updated via
\beq
  \hat{L}_k(\theta_i) \,=\, \hat{L}_{k-1}(\theta_i) \,+\, \log\Vert\hat{\bt}_k^{\theta_i}\Vert_1 \,,
  \quad i=1,\ldots,M\,,
\label{eq-MLdisc1}
\eeq
with $\hat{L}_0(\theta_i) = 0$, for $i=1,\ldots,M$, and the ML
estimator for both parameter and state at time $k$ can be directly
obtained by running $M$ particle filters in
parallel, each tuned to a candidate parameter $\theta_i$, for
$i=1,\ldots,M$: 
\beq
 \hat{\theta}^{\rm ML}_k \,=\, \argmax_{\theta \in \{\theta_1,\ldots,\theta_M\!\}}
  \hat{L}_k(\theta)\,,
\label{eq-MLdisc2}
\eeq
\beq
  \hXk^{\rm ML} =\, \hXk^{\rm MS}(\hat{\theta}^{\rm ML}_k)\,,
\label{eq-MLdisc3}
\eeq
for $k=1,2,\ldots$ 

\begin{figure*}[ht!]
\center
\includegraphics[width=133mm]{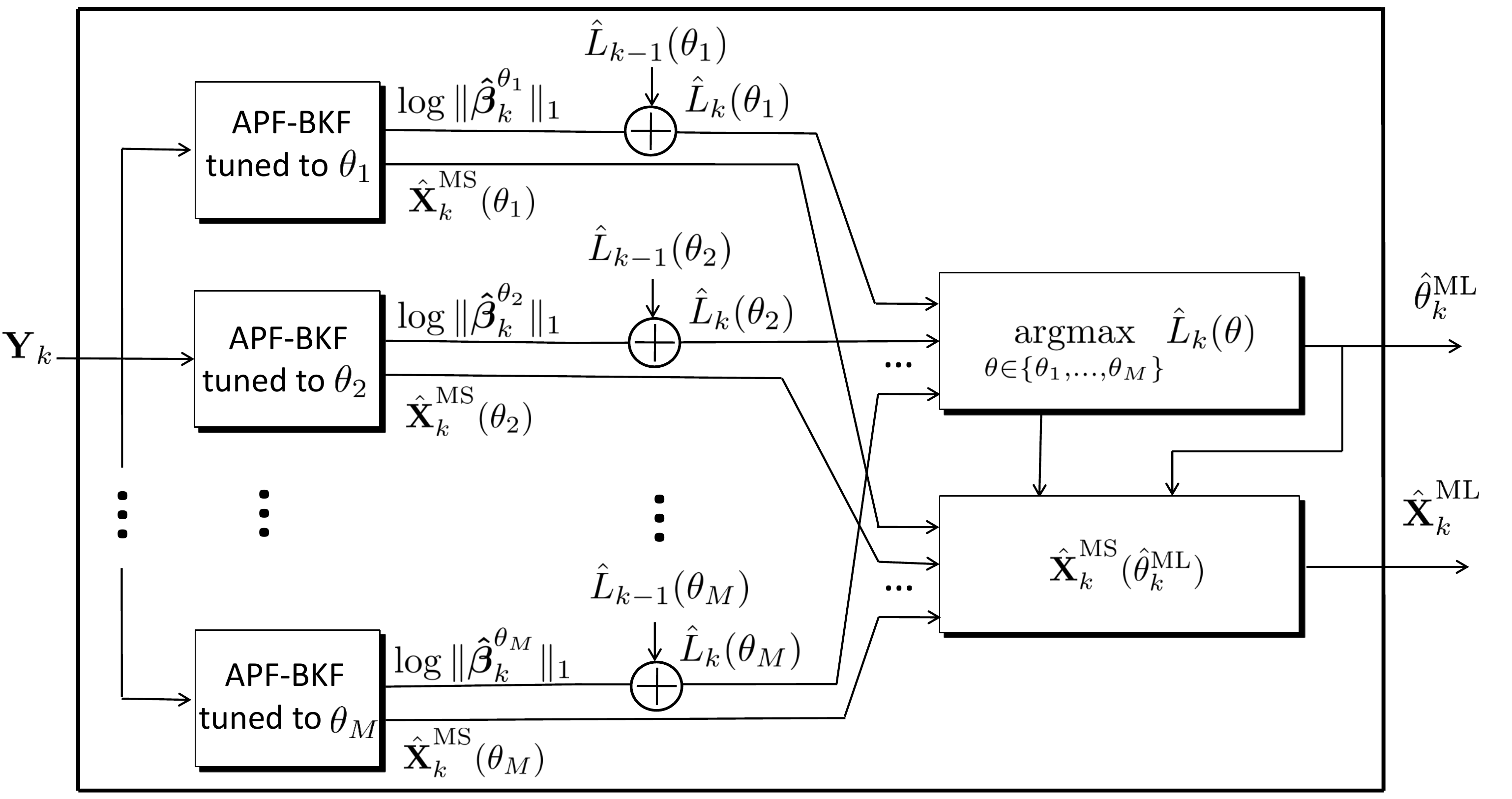}
\caption{Schematic diagram of particle-filter implementation of the 
 discrete-parameter ML adaptive Boolean Kalman Filter.}
\label{fig:MMAE}
\end{figure*}

The computation in (\ref{eq-MLdisc1})--(\ref{eq-MLdisc3}) is
parallelized, on-line, and entirely recursive: as a new observation at
time $k+1$ arrives, the ML estimator can be updated easily without
restarting the computation from the beginning. The procedure is
summarized in Figure~\ref{fig:MMAE} and Algorithm~\ref{alg:Discrete}.

\begin{algorithm}
\caption{APF-DPMLA-BKF: APF implementation of the
  discrete-parameter ML Adaptive BKF}
\label{alg:Discrete}
\begin{algorithmic}[1]
\small
\State $\hat{L}_0(\theta_i) \,= \,1$, for $i=1,\ldots,M$.\vspace{0.5ex}
\For{$k =1,2,\ldots,$} 
\State  Run $M$ APF-BKFs tuned to $\theta_1,\ldots\theta_M$:
\State $\hat{L}_k(\theta_i) \,=\, \hat{L}_{k-1}(\theta_i)+\log\Vert\hat{\bt}_k^{\theta_i}\Vert_1\,$, $i=1,\ldots,M$.\vspace{0.8ex}
\State $\hat{\theta}^{\rm ML}_k \,=\, \arg\max_{\theta_1,\ldots,\theta_M}  \hat{L}_k(\theta_i)$.\vspace{0.8ex}
\State $\hXk^{\rm ML} \,=\, \hXk^{\rm MS}(\hat{\theta}^{\rm ML}_k)$.\vspace{0.8ex}
\EndFor
\end{algorithmic}
\end{algorithm}

\subsection{APF Implementation of the Continuous-Parameter ML Adaptive BKS
  (APF-CPMLA-BKS)} 
\label{sec:CPE}

Here, $\Theta \subseteq R^m$, and the approach developed in the last
subsection for discrete parameter spaces is not directly
applicable. There are two options: 1) discretize the parameters using
a suitable quantization grid and apply an approach similar to the one
in the last section; 2) attempt to obtain a good approximation of the
MLE in the continuous parameter space directly. In this section, we
describe how to implement the second option using the
expectation-maximization (EM) algorithm for a particle filter
implementation of a fixed-interval Boolean Kalman Smoother.

In our case, maximum likelihood estimation attempts to
find the value of $\theta$ that maximizes the ``incomplete''
log-likelihood function $L_k(\theta) = \log\,p_{\theta}(\YoT)$.
The EM algorithm considers instead the ``complete''
log-likelihood function $\log p_{\theta}(\XoT,\YoT)$,
which includes the unknown state sequence, the assumption being that
maximising the complete log-likelihood is easier than maximising the
incomplete one (the reader is referred to~\cite{ImanBrag:16b} for more
information on the EM algorithm for POBDS). 

The EM algorithm obtains a sequence of parameter estimates $\{\thetak;
n=0,1,\ldots\}$. Given the current estimate $\thetak$, the algorithm
obtain the next estimate $\thetakp$ in the sequence by computing (E-step) the
function (see~\cite{ImanBrag:16b}):
\beq\label{eq:Qf}
\bal
  Q(\theta,\thetak)
  &\,=\,\sum_{\XoT} \log p_{\theta}(\XoT,\YoT)\, p_{\thetak}(\XoT \mid \YoT) \\[0.5ex]
  & \,=\,I_1(\theta,\thetak) + I_2(\theta,\thetak) +
I_3(\theta,\thetak)\,,
\eal
\eeq
where
\vspace{-2ex} 
\beq\label{eq:EM5}
\bal
I_1(\theta,\thetak) &=\sum_{i=1}^{2^d}\,\log
P_{\theta}(\Xz=\x^i)\,P_{\thetak}(\Xz=\x^i\mid\YoT),
\eal
\eeq
\vspace{-2ex} 
\beq\label{eq:EM6}
\bal
& I_2(\theta,\thetak) \,=\, \sum_{s=1}^T \sum_{i=1}^{2^d}\sum_{j=1}^{2^d}\,\log P_\theta(\Xs=\x^i\mid\Xsm=\x^j) \\
 &\qquad\qquad\qquad \times\, P_{\thetak}(\Xs=\x^i,\Xsm=\x^j\mid \YoT)\,,
\eal
\eeq
\vspace{-2ex} 
\beq\label{eq:EM7}
\bal
I_3(\theta,\thetak) & \,=\, \sum_{s=1}^T \sum_{i=1}^{2^d} \log
p_\theta(\Ys \mid \Xs=\x^i)\\[-1ex]
& \qquad\qquad\qquad \times\, P_{\thetak}(\Xs=\x^i \mid \YoT)\,,\\
\eal
\eeq
and then maximizing (M-step) this function:
\beq
  \thetakp \,=\, \argmax_{\theta} Q(\theta,\thetak)\,.
\eeq

In \cite{ImanBrag:16b} this computation is carried out exactly. For
large systems, this is impractical, for the following reasons:

\begin{enumerate}
\item The E-Step (computing the $Q$ function) requires
  performing a Boolean Kalman Smoother, which is too expensive computationally.

\item The transition matrix and filtered and smoothed posterior
  probability vectors at all time steps must be stored, demanding
  large amounts of memory.

\item In certain cases, such as when $\f_k$ and $\h_k$ are linear in
  the parameter vector $\theta$, it is possible to maximize
  ${Q}(\theta , \thetak)$ using closed-form expressions
  (e.g.~\cite{barembruch2009approximate}). However, in general, one
  needs to resort to gradient-based optimization methods in the
  M-step. This requires evaluating ${Q}(\theta , \thetak)$ and
  computing its derivatives, which is analytically intractable.
\end{enumerate}

To address all these issues, we develop our EM algorithm based on the
\textit{Forward Filter Backward Simulation}~\cite{godsill2012monte}
method. This method tries to capture the most probable state
trajectories and use them to find the smoothing particles. The method
contains two main steps: 1) Forward Step: the APF-BKF algorithm is
employed to obtain the particles and their weights from time 0 to $T$
($\{\x_{1:T,i},W_{1:T,i}\}_{i=1}^N$). 2) Backward Step: the backward
simulation procedure, which is explained in detail in the sequel,
computes $N$ trajectories
$\{\tx_{0:T,i}\}_{i=1}^N\,\sim\,P(\X_{0:T}\mid\Y_{1:T})$, where
\beq\label{eq-joint}
\bal
&P(\XoT\mid\YoT)\,\\
&=\, P(\X_{T}\mid\Y_{1:T})\,\prod_{s=0}^{T-1} P(\X_{s}\mid\X_{s+1:T},\Y_{1:T})\\
&=\, P(\X_{T}\mid\Y_{1:T})\,\prod_{s=0}^{T-1} P(\X_{s}\mid\X_{s+1},\Y_{1:s})\,.
\eal
\eeq
Based on equation~(\ref{eq-joint}), smoothed particles can be obtained
using the FFBSi method, by means of the following backward procedure:
\beq\label{eq:FF1}
\bal
\tilde{\x}_{T,i}\,&\sim\,P(\X_{T}\mid\Y_{1:T})\,,\\[1ex]
\tilde{\x}_{s,i}\,&\sim\,P(\X_{s}\mid\tilde{\x}_{s+1,i},\Y_{1:T})\,,
\eal
\eeq
for $i=1,\ldots,N$ and $s=T-1,\ldots,0$, where $\{\tx_{s,i}\}_{i=1}^N$
are the smoothed particles at time step $s$.

The backward process starts by 
resampling $N$ particles
$\{\tx_{T,i}\}_{i=1}^N$ from the unique forward particles $\{\x^u_{T,i}\}_{i=1}^{F_T}$ at time step
$T$ using the forward weights
$\{W_{T,i}^u\}_{i=1}^{F_T}$ as:
\beq
\bal
 \{\eta_T(i)\}_{i=1}^N \,&\sim \,\text{Cat}\,(\{W^u_{T,j}\}_{j=1}^{F_T})\,,\\[1ex]
 \tx_{T,i}\,&=\,\x^u_{T,\eta_T(i)}, \:\:i=1,\ldots,N\,.
\eal
\eeq
Now, to obtain $N$ smoothed particles at time step
$s<T$, let $\{\tx_{s+1,i}\}_{i=1}^{N}$ be the smoothed particles at
time $s+1$, and let
\beq\label{eq:uni}
\{\tx^u_{s+1,j},\xi_{s+1}^j\}_{j=1}^{S_{s+1}} \xleftarrow{\text{Unique}} \{\tx_{s+1,i}\}_{i=1}^{N}
\eeq 
where $S_{s+1}$ specifies the number of unique smoothed particles at time $T$, and
$\xi_{s+1}^j$ contains the indexes of the $j$-th unique smoothed particles
before shrinkage and reordering. Notice that $N=\sum_{j=1}^{S_{s+1}} |\xi_T^j|$.
For the $j$-th unique smoothed particle at time step $s+1$, one can use the fact that $P(\X_s\mid\X_{s+1},\Y_{1:s})\propto P(\X_{s+1}\mid\X_s)\,P(\X_s\mid\Y_{1:s})$ to
compute the following weights
\beq\label{eq:wm}
D_{s,i}^j\,=\,W^u_{s,i}\,P(\tx^u_{s+1,j}\mid\x^u_{s,i})\,, 
\eeq
for $i=1,\ldots,F_s$, and draw $|\xi_{s+1}^j|$
particles as: 
\beq
\bal
\{\eta_s(t)\}_{t=1}^{|\xi_{s+1}^j|}\,&\sim\, {\rm Cat}(\{{D}^j_{s,i}\}_{i=1}^{F_s})\,, \\[0.5ex]
\tx_{s,\xi^j_{s+1}(t)} \,&=\, \x^u_{s,\eta_s(t)}, \:\:\text{ for } t=1,\ldots,|\xi_{s+1}^j|.
\eal
\eeq

Repeating the above process for $j=1,\ldots,S_T$ and $s=T-1,\ldots,0$
results in $N$ trajectories from the joint smoothed distribution
$\{\tx_{0:T,i}\}_{i=1}^N$.  Notice that the computational complexity
of this method is of order $O(S_s\times F_s)$ at time step $s$, which
can be much smaller than the computational complexity of the optimal
smoother which is $O(N\times N)$.

Given the $N$ trajectories $\{\tx_{0:T,i}\}_{i=1}^N$ obtained by running the Forward Filter Backward Simulation tuned to parameter $\thetak$, equations~(\ref{eq:EM5})-(\ref{eq:EM7}) can be approximated as: 
\begin{align}
\hat{I}_1(\theta,\thetak) \,&=\,\frac{1}{N}\sum_{i=1}^{N}\,\log P_{\theta}(\tilde{\x}_{0,i})\,,\\
 \hat{I}_2(\theta,\thetak) \,&=\, \frac{1}{N} \sum_{s=1}^T \sum_{i=1}^{N} \log P_\theta(\tilde{\x}_{s,i}\mid\tilde{\x}_{s-1,i})\,,\\
 \hat{I}_3(\theta,\thetak) \,&=\,\sum_{s=1}^T \sum_{i=1}^{N} \log p_\theta(\Y_s \mid \tilde{\x}_{s,i})\,.
\end{align}
Thus, one can approximate the $Q$ function in equation~(\ref{eq:Qf}) as:
\beq\label{eq:EMp8}
\bal
\hat{Q}&(\theta,\theta^{(n)})\, =\, \frac{1}{N}\,\sum_{i=1}^{N} \bigg[\log P_{\theta}(\tilde{\x}_{0,i})\\
& + \sum_{s=1}^T \log P_\theta(\tilde{\x}_{s,i}\mid\tilde{\x}_{s-1,i})
+\sum_{s=1}^T\log\,p_\theta(\Y_s \mid \tilde{\x}_{s,i})\bigg].
\eal
\eeq

In~\cite{wills2013identification}, similar equations are derived for
the Hammerstein-Wiener model structure. The computational complexity
of evaluating the $\hat{Q}$ function is only of order $O(Nk)$, which
results in large savings in the computation of the gradient of
$\hat{Q}$ in the M-Step of the EM algorithm. In Section~\ref{sec:GRN},
expressions for the gradients are given in the special case where the
observations consist of RNA sequencing data. Finally, as regards to
memory, the only values that must be stored from the E-Step to be used
in the M-Step are the $N$ smoothed trajectories (storing filter
weights or particles is not necessary). In Section~\ref{sec:NE}, the
effect of the value of $N$ on performance will be discussed.

\begin{figure*}[ht!]
\center
\includegraphics[width=133mm]{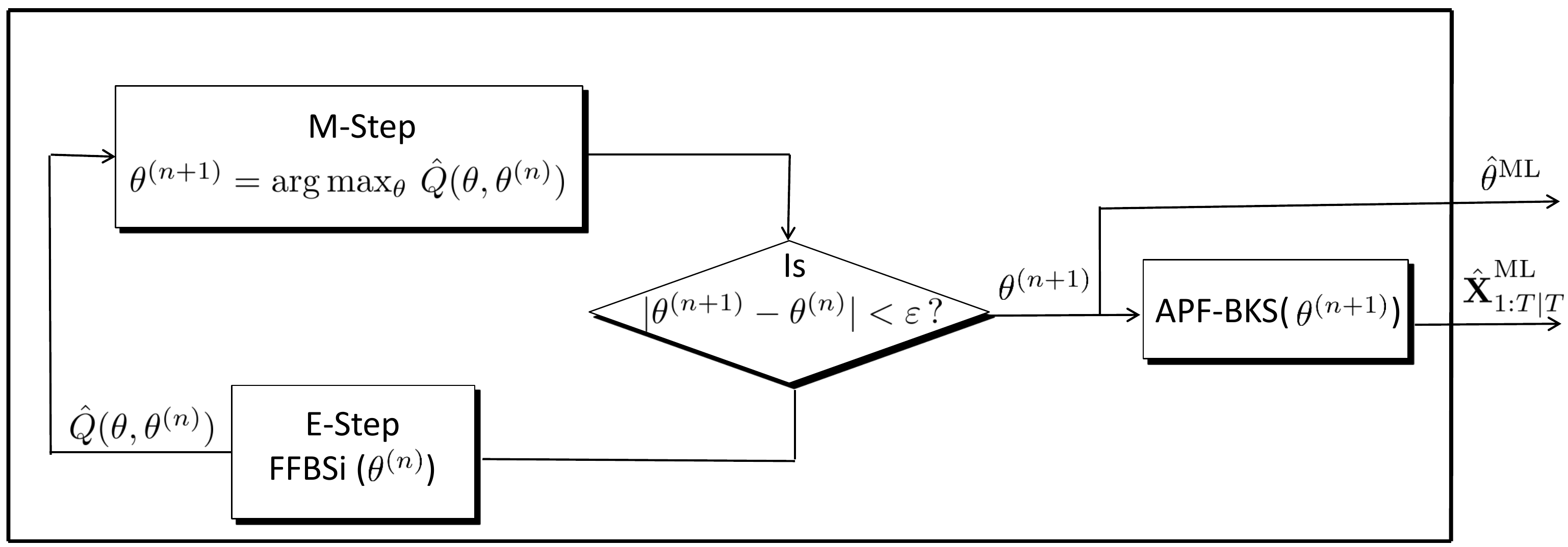}
\caption{Schematic diagram of particle-filter implementation of the 
 continuous-parameter ML adaptive BKS.}
\label{fig:EM}
\end{figure*}

The steps of the EM adaptive filter are as follows.  Initially, $N$
smoothed trajectories are obtained using the developed FFBSi method
tuned to a initial parameter guess $\theta^{(0)}$ to compute
$\hat{Q}(\theta,\theta^{(0)})$ (E-Step). The aforementioned
gradient-descent procedure is applied to find the best parameter
$\theta^{(1)}$ that maximizes $\hat{Q}(\theta,\theta^{(0)})$ with
$\theta^{(0)}$ fixed (M-Step). The obtained parameter vector is set as
the parameter for the particle smoother for the next run, and the
process continues until there is no significant change in parameter
estimates between two consecutive steps, yielding the final parameter
estimate $\theta^{\rm ML}$.  Then the smoothed state estimates can be
obtained by performing an APF-BKS tuned to parameter
$\theta^{\rm ML}$. The procedure is summarized in Figure~\ref{fig:EM}
and Algorithm~\ref{alg:EM}.

\begin{algorithm}
\caption{APF-CPMLA-BKS: APF implementation of the
  continuous-parameter ML Adaptive BKS.}
\label{alg:EM}
\begin{algorithmic}[1]
 \small
\State Specify $\theta^{(0)}$ (initial guess) and
  tolerance $\varepsilon>0$.\vspace{0.8ex}
\State $n \leftarrow -1$. \vspace{0.8ex}
\Repeat \vspace{0.8ex}
\State $n \leftarrow n+1$\vspace{0.8ex}
\State $\{\x_{0:T,i},W_{0:T,i}\}_{i=1}^N\,\leftarrow\,$Run APF-BKF tuned to $\thetak$.\vspace{0.8ex}
\State $\{\x^u_{k,j},W^u_{k,j}\}_{j=1}^{F_{k}}\!\xleftarrow{\text{Unique}}\!\{\x_{k,i},W_{k,i}\}_{i=1}^{N},\,k=0,\ldots,T$.\vspace{1ex}
\State Sample $\{\eta_T(i)\}_{i=1}^N \,\sim \,\text{Cat}\,(\{W^u_{T,j}\}_{j=1}^{F_T})$.\vspace{0.8ex}
\State Set $\tx_{T,i}=\x^u_{T,\eta_T(i)}$, for $i=1,\ldots,N$.\vspace{0.8ex}
\For{$s =T-1$ to $0$} \vspace{0.8ex}
\State $\{\tx^u_{s+1,j},\xi_{s+1}^j\}_{j=1}^{S_{s+1}} \xleftarrow{\text{Unique}} \{\tx_{s+1,i}\}_{i=1}^{N}$\vspace{0.8ex}
 \For{$j =1$ to $S_{s+1}$}\vspace{0.8ex}
   \State $D_{s,i}^j\,=\,{W^u_{s,i}\,P(\tx^u_{s+1,j}\mid\x^u_{s,i})} ,\, i=1,\ldots,F_s$.\vspace{0.8ex} 
\State $\{\eta_s(t)\}_{t=1}^{|\xi_{s+1}^j|}\sim \text{Cat}\,(\{D^j_{s,i}\}_{i=1}^{F_s}).$\vspace{0.8ex}
\State $\tx_{s,\xi^j_{s+1}(t)} \,=\, \x^u_{s,\eta_s(t)}, \text{ for } t=1,\ldots,|\xi_{s+1}^j|.$\vspace{0.8ex}
\EndFor\vspace{0.8ex}
\EndFor\vspace{0.8ex}
\State Find $\hat{Q}(\theta,\theta^{(n)})$ using equation~(\ref{eq:EMp8}).\vspace{0.8ex}
\State Find $\theta^{(n+1)}\, =\, \argmax_{\theta} \hat{Q}\left(\theta,\theta^{(n)}\right)$.\vspace{0.8ex}
\Until{$\vert \theta^{(n+1)}-\theta^{(n)}\vert> \varepsilon$}\vspace{0.8ex}
\State $\hat{\theta}^{\text{ML}}=\thetakp$. \vspace{0.8ex}
\State $\hat{\X}_{1:T|T}^{\rm MS}(\hat{\theta}^{\text{ML}})\,\leftarrow\,$Run APF-BKS tuned to $\hat{\theta}^{\text{ML}}$.\vspace{0.8ex}
\State $\hat{\X}_{1:T|T}^{\rm ML} = \hat{\X}_{1:T|T}^{\rm MS}(\hat{\theta}^{\text{ML}})$.
\end{algorithmic}
\end{algorithm}

\section{Gene Regulatory Network and
  RNA-Seq Measurement Models}
\label{sec:GRN}

The algorithms developed in the previous section apply to the general
partially-observed Boolean dynamical systems signal model in
(\ref{eq-sgnmodel}). In this section, we describe a specific instance
of that model, which allows the application of the methodology to
Boolean gene regulatory networks observed through next-generation
sequencing measurements. The gene regulatory network model corresponds
to the state model in the general POBDS model, whereas the RNA-seq
measurement model corresponds to the observation model.
 
\subsection{Gene Regulatory Network Model}
\label{sec:GRNa}

\begin{figure}[ht!]
\centering
\includegraphics[width=70mm]{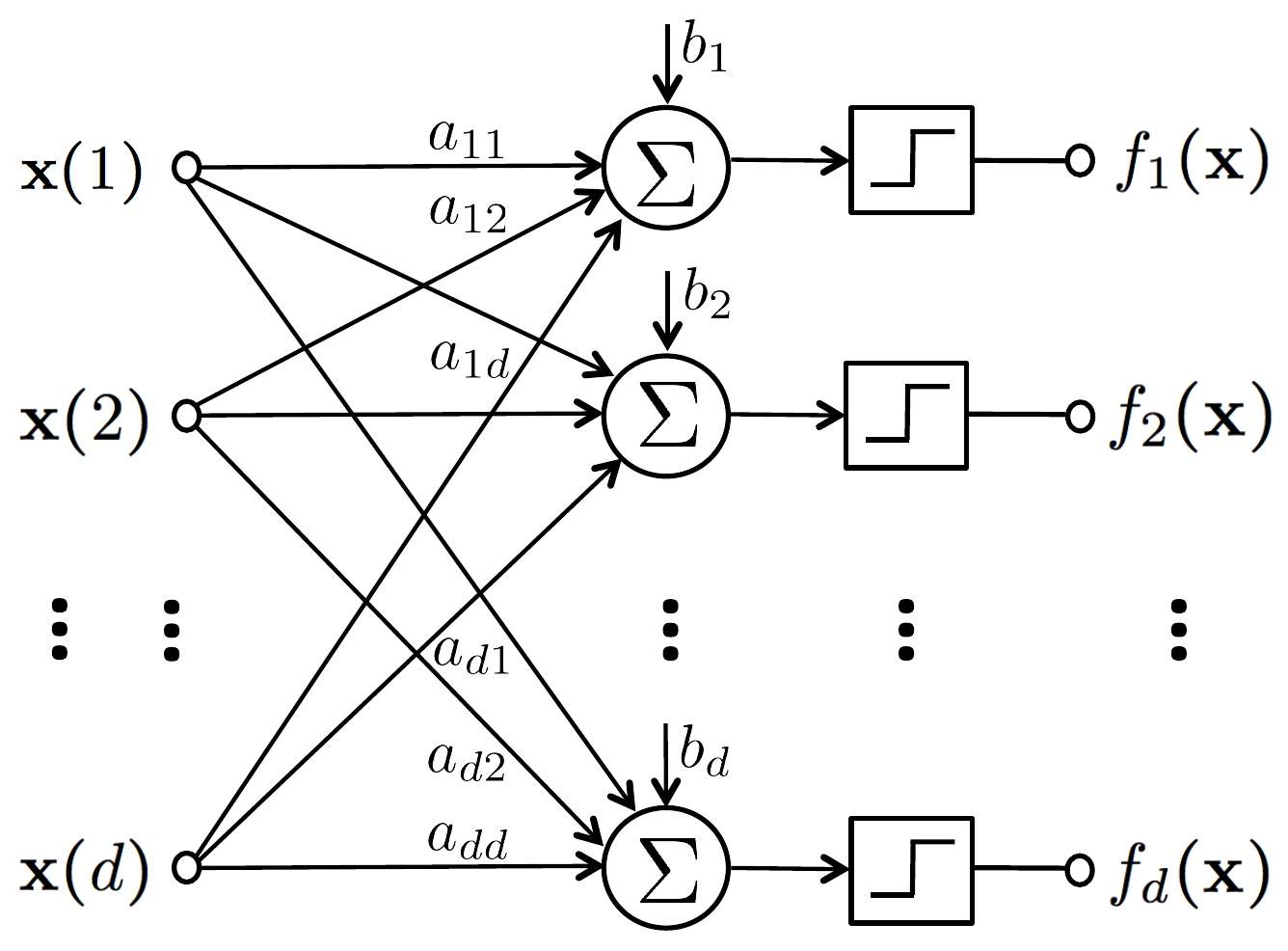}
\caption{Gene regulatory network model.}
\label{fig:GRN}
\end{figure}

This model is motivated by gene pathway diagrams commonly encountered in
biomedical research. The network function in (\ref{eq-sgnmodel}) is
assumed to be time-invariant and expressed as $\f = (f_1,\ldots,f_d)$, where each
component $f_i: \{0,1\}^{d} \rightarrow \{0,1\}$ is a Boolean
function given~by
\beq\label{eq:BN3}
f_i(\x) \,=\, 
\begin{cases}
1, & \sum_{j=1}^d a_{ij} \x(j) + b_i>0 \,, \\
0, & \sum_{j=1}^d a_{ij} \x(j) + b_i\leq 0 \,, 
\end{cases}
\eeq
where $a_{ij}$ and $b_i$ are system parameters. The former can take
three values: $a_{ij}=+1$ if there is positive regulation (activation)
from gene $j$ to gene $i$; $a_{ij}=-1$ if there is negative regulation
(inhibition) from gene $j$ to gene~$i$; and $a_{ij}=0$ if gene $j$ is
not an input to gene $i$.  The latter specifies regulation {\em
  biases} and can take two values: $b_i = +1/2$ or $b_i = -1/2$.  The
network function is depicted in Figure~\ref{fig:GRN}, where the
threshold units are step functions that output 1 if the input is
nonnegative, and 0, otherwise.

The process noise $\nk$ in (\ref{eq-sgnmodel}) is assumed to have
independent components with $P(\nk(i)=1)= p$, for $i= 1,\ldots,d$, and
$k=1,2,\ldots$ The noise parameter $0 \leq p \leq 0.5$ gives the
amount of ``perturbation" to the Boolean state process; the closer it
is to $p = 0.5$, the more chaotic the system will be, while a value of
$p$ close to zero means that the state trajectories are nearly
deterministic, being governed tightly by the network function.


\subsection{RNA-Seq Measurement Model}
\label{sec-obs}

Next-generation sequencing (NGS) technologies are able to sequence
millions of short DNA fragments in parallel; the length and number of
reads vary with the specific technology \cite{ghaffari2013modeling}.
The application of NGS technology to transcriptional profiling is
called RNA-seq, which records how frequently each transcript is
represented in a sequence sample \cite{mortazavi2008mapping}. RNA-seq
is a probe-free approach that can capture any relevant transcript
present in a sample, without the need of prior knowledge about the
target sequence.

Let $\Yk = (\Yk(1),\ldots,\Yk(d))$ be a vector containing the RNA-seq
data at time $k$, for $k = 1,2,\ldots$ such that $\Yk(j)$ is the read
count corresponding to transcript $j$ in a single-lane platform, for
$j = 1,\ldots,d$. 
We assume conditional
independence of the transcript counts given the state,
\beq
\bal
& P\left(\Yk = \y \mid \Xk = \x\right)\\
& \quad \,=\, \prod_{j=1}^d P(\Yk(j) =
\y(j) \mid \Xk(j)=\x(j))\,,
\eal
\label{eq:NGS0}
\eeq
and adopt the negative binomial model for each count,
\beq
\label{eq:NGS1}
\bal
P(&\Yk(j) = \y(j) \mid \Xk(j)=\x(j))\, = \\
 &\frac{\Gamma(\y(j)+\phi_j)}{\y(j)! \,\Gamma(\phi_j)}\, \left(\frac{\lambda_{j}}{\lambda_{j}+\phi_j}\right)^{\y(j)}\!\left(\frac{\phi_j}{\lambda_{j}+\phi_j}\right)^{\phi_j}\!,
\eal
\eeq
where $\Gamma$ denotes the Gamma function, and $\phi_j,\lambda_{j}>0$
are the real-valued inverse dispersion parameter and mean read count
of transcript $j$, respectively, for $j=1,\ldots,d$. The inverse
dispersion parameter models observation noise; the smaller it is, the
more variable the measurements are.

Now, recall that, according to the Boolean state model,
there are two possible states for the abundance of transcript $j$:
high, if $\x(j) = 1$, and low, if $\x(j) = 0$. Accordingly, we
model the parameter $\lambda_{j}$ in log-space~as:
\beq
\log \lambda_{j}\,=\,\log s \,+\, \mu \,+\, \delta_j\,\x(j)\,,
\label{eq:NGS2}
\eeq
where the parameter $s$ is the {\em sequencing depth} (which is
instrument-dependent), $\mu>0$ is the baseline level of expression
in the inactivated transcriptional state, and $\delta_j > 0$ expresses
the effect on the observed RNA-seq read count as gene $j$ goes from
the inactivated to the activated state, for $j=1,\ldots,d$.

Based on equations~(\ref{eq:NGS0})--(\ref{eq:NGS2}), given the particles $\tx_{k,i}$, one can compute $P(\Y_k\mid\X_k=\tx_{k,i})$ as :
\beq
\bal
&P\left(\Yk = \y \mid \Xk = \tx_{k,i}\right)\\
&=\,\prod_{j=1}^d \,\left[\frac{\Gamma(\y(j)+\phi_j)}{\y(j)!
  \,\Gamma(\phi_j)}\, \left(\frac{s\,\exp(\mu+\delta_j\,\tx_{k,i}(j))}{s\,\exp(\mu+\delta_j\,\tx_{k,i}(j))+\phi_j}\right)^{\y(j)} \right.\\
& \quad\qquad\qquad\qquad \left. \times \,\left(\frac{\phi_j}{s\,\exp(\mu+\delta_j\,\tx_{k,i}(j))+\phi_j}\right)^{\phi_j}\right]\,.
\eal\label{eq:GRN5}
\eeq

The RNA-seq measurement model parameters are thus the sequencing depth
$s$, 
the baseline expression level $\mu$, the transcript-dependent differential
expression levels $\delta_j$, for $j=1,\ldots,d$, and the
transcript-dependent inverse dispersion parameters $\phi_j$, for
$j=1,\ldots,d$. These are all continuous parameters.

\begin{table*}[ht!]
\caption{Derivatives of $Q(\theta,\theta^{(n)})$ with respect to different parameters needed in Numerical Experiment.}
\centering
\begin{tabular}{l*{2}{cc}r}
\midrule
Parameter & derivation \\
\midrule
$p$ &\(\displaystyle
\frac{1}{N}\sum_{s=1}^T \sum_{i=1}^{N} \left( \frac{\Vert\tx_{s,i}\,\oplus\,\f(\tx_{s-1,i})\Vert_1- d\,p}{p\,(1-p)}\right)
\)\\
\midrule
$s$ & {\(\displaystyle\frac{1}{Ns} \,\sum_{s=1}^T \,\sum_{i=1}^{N}
\sum_{j=1}^d\,\left(\frac{\phi_j\,\left(\Y_s(j)-s\,\exp(\mu+\delta_j \,\tx_{s,i}(j))\right)}{\phi_j+ s\,\exp(\mu+\delta_j \,\tx_{s,i}(j))}\right)\)}\\
\midrule
$\mu$ & {\(\displaystyle\frac{1}{N} \,\sum_{s=1}^T \,\sum_{i=1}^{N}
\sum_{j=1}^d\,\left(\frac{\phi_j\,\left(\Y_s(j)-s\,\exp(\mu+\delta_j \,\tx_{s,i}(j))\right)}{\phi_j+ s\,\exp(\mu+\delta_j \,\tx_{s,i}(j))}\right)\)} \\
\midrule
$\delta_j$ & \(\displaystyle \frac{1}{N}  \sum_{s=1}^T \,\sum_{i=1}^{N} \left(\frac{\tx_{s,i}(j)\,\phi_j\,\left(\Y_s(j)-s\,\exp(\mu+\delta_j \,\tx_{s,i}(j))\right)}{\phi_j+s\,\exp(\mu+\delta_j \,\tx_{s,i}(j))}\right)\) \\
\midrule
$\phi_j$ & \(\displaystyle \frac{1}{N} \sum_{s=1}^T \,\sum_{i=1}^{N}
 \left(\frac{\Gamma'(\Y_s(j)+\phi_j)}{\Gamma(\Y_s(j))+\phi_j)}-\frac{\Gamma'(\phi_j)}{\Gamma(\phi_j)}+\frac{s\,\exp(\mu+\delta_j \,\tx_{s,i}(j))-\Y_s(j)\phi_j}{\phi_j(s\,\exp(\mu+\delta_j \,\tx_{s,i}(j))+\phi_j)}+\log\frac{\phi_j}{s\,\exp(\mu+\delta_j \,\tx_{s,i}(j))+\phi_j}\right)\) \\
\midrule
\end{tabular}
\label{table:grad}
\end{table*}

\section{Numerical Experiments}
\label{sec:NE}
\begin{figure}[ht!]
\centering
\includegraphics[width=80mm]{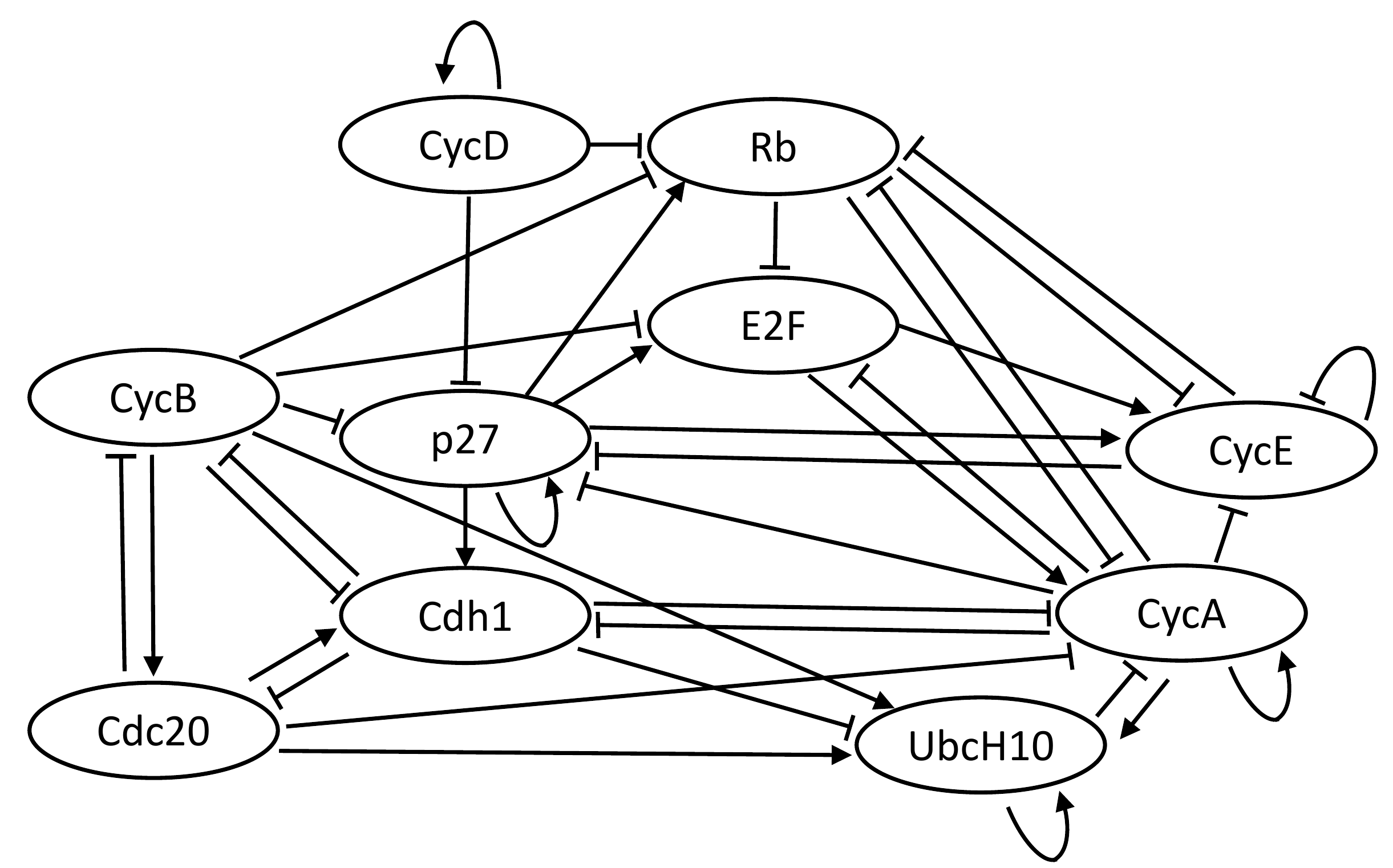}
\caption{Pathway diagram for the cell-cycle network.}
 \label{fig:cell}
\end{figure}

In this section, we carry out detailed numerical experiments to assess
the performance of the developed particle-based methods. We base our experiments on
the well-known {\em Mammalian Cell-Cycle network}~\cite{faure2006dynamical}. The pathway diagram for this network is presented in
Figure~\ref{fig:cell}.
The state vector is $\x =$
(\textit{CycD}, \textit{Rb}, \textit{p27}, \textit{E2F}, \textit{CycE}, \textit{CycA}, \textit{Cdc20}, \textit{Cdh1}, \textit{UbcH10}, \textit{CycB}). The gene
interaction parameters $a_{ij}$ can be read off Figure~\ref{fig:cell} easily. As an example, \textit{Rb} is activated by \textit{p27}, and is inactivated by \textit{CycD}, \textit{CycE}, \textit{CycA},  \textit{CycB}. These interactions can be expressed in terms of interaction parameters as: $a_{21}=-1$, $a_{22}=0$, $a_{23}=+1$, $a_{24}=0$, $a_{25}=-1$, $a_{26}=-1$, $a_{27}=0$, $a_{28}=0$, $a_{29}=0$ and $a_{2\,10}=-1$.

In all numerical experiments to follow, we assume the same fixed set
of ``true'' values for the system parameters, summarized in
Table~\ref{table:para}.
\begin{table}[ht!]
\centering
\caption{Parameter values for numerical experiments using the Cell-Cycle gene regulatory network.}
\begin{tabular}{l*{1}{cc}r}
\midrule
\textbf{Parameter} & \textbf{Value}\\
\midrule
\text{Length of time series $T$}  & 50, 100 \\
\midrule
\text{Number of genes $d$}  & 10 \\
\midrule
\text{Initial distribution $P(\X_0=\x^i), i=1:2^{10}$}  & $1/2^{10}$ \\
\midrule
\text{Number of particles $N$}  & 200, 1000, 5000 \\
\midrule
\text{Bias $b_i, i=1,\ldots,10$}  & -1/2 \\
\midrule
Transition noise intensity $p$ & 0.01, 0.05 \\
\midrule  
Sequencing depth $s$ &\hspace{-3em} 1.02 (1K-50K reads) \\
\midrule
\text{Baseline expression $\mu$} & 0.1\\
\midrule
\text{Differential expression $\delta_i,\, i=1,\ldots,10$ } & 2\\ 
\midrule
\text{Inverse dispersion $\phi_i,\, i=1,\ldots,10$} & 1, 5 \\[2pt]
\midrule
APF-CPMLA-BKS stopping criterion $\varepsilon$ & $10^{-4}$ \\[2pt]
 \midrule
\end{tabular}
\label{table:para}
\end{table}

\subsection{Experiment 1: State Estimation}

In this section, the state estimation performance of the APF-BKF and
APF-BKS is compared to that of the exact BKF and BKS, respectively.
Given that the cell-cycle network comprises 10 genes, the size of the
transition and update matrices required by both the BKF and BKS is
$2^{10}\times 2^{10}$. As a result, the computational cost of the BKF
and BKS is high.  Table~\ref{table:PF} shows the average rate of
correct state estimation over 1000 independent runs for a time series
with length 100.

\begin{table}[ht!]
\caption{Experiment 1: Average rates of correct state estimation over 1000 independent runs for a time series
with length 100.}
\centering
\begin{tabular}{ccccccc}
\midrule
\small{$p$}  & $\phi$   & $N$  & BKF & APF-BKF & BKS &APF-BKS \\
\midrule
\multirow{7}{*}{0.01} & \multirow{3}{*} {5} & 200 &  \multirow{3}{*}{93.9} & 85.4 & \multirow{3}{*}{96.6} & 88.1\\[2pt]
& & 1000 & & 92.1 &  & 95.0 \\
& & 5000 & & 93.2 &   & 95.7 \\
\cmidrule{2-7}
&\multirow{3}{*}{1} & 200 &  \multirow{3}{*}{83.8}& 74.6  &\multirow{3}{*}{90.7} &  80.4   \\[2pt]
& &  1000 & & 80.6&   & 88.3   \\
& &  5000 & & 82.1&   & 89.8   \\
\midrule
\multirow{7}{*}{0.05} & \multirow{3}{*}{5} & 200 &  \multirow{3}{*}{82.9} & 75.0 & \multirow{3}{*}{93.4} & 82.3  \\[2pt]
&  & 1000 &&80.3   & & 91.3 \\
&  & 5000 &&81.9   & & 92.6 \\
\cmidrule{2-7}
&\multirow{3}{*} {1} & 200& \multirow{3}{*}{58.5} & 50.1   &\multirow{3}{*}{70.8} & 62.3   \\[2pt]
& & 1000 & & 55.1 & & 68.2   \\
& & 5000& & 56.9 & & 69.9   \\
\midrule
\end{tabular}
\label{table:PF}
\end{table}

As expected, the performance of both the AFP-BKF and APF-BKS is higher
for large number of particles. However, the improvement is
significantly larger by moving from 200 to 1000 particles in
comparison to moving from 1000 to 5000 particles. One can also see the
reduction in performance of all filters and smoothers as process noise
or dispersion in measurements increases, which both make the
estimation process more challenging.  Also as expected, the BKS and
APF-BKS outperform the BKF and APF-BKF, respectively, due to the
availability of more data for estimation.

\begin{figure}[ht!]
\centering
\includegraphics[width=85mm]{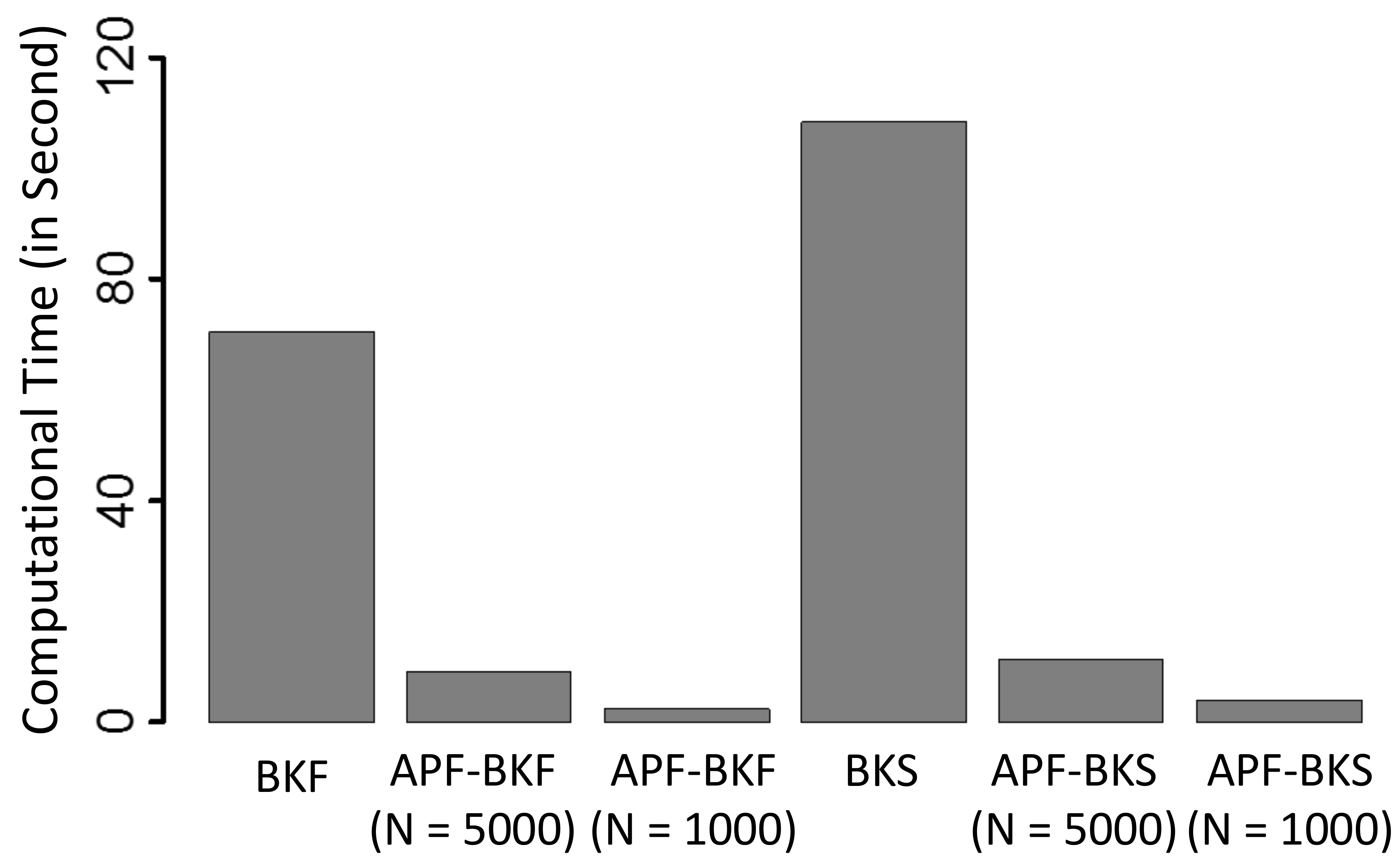}
\caption{Experiment 1: Average computation time (in seconds) for the exact and
  partice-based algorithms.}
\label{fig:time}
\end{figure}

The average time to run the various
algorithms for time series of length 100 and different number of
particles is displayed in Figure~\ref{fig:time}.  Here, $p=0.05$ and
$\phi=5$.
The average computational time is measured on a PC with an Intel Core i7-4790
CPU@3.60 GHz clock and 16 GB of RAM.  The results show the very large
computational savings afforded by the APF-BKF and APF-BKS in
comparison to the exact algorithms.

\subsection{Experiment 2: Incomplete Network Topology}

In this experiment, we assume that the interaction between genes
\textit{Rb} and \textit{E2F}, or equivalently the gene interaction
parameter $a_{42}$, is unknown, and all other parameters are
known. Since this is a discrete parameter estimation problem, the
APF-DPMLA-BKF is run, which in this case consists of three APF-BKFs
running in parallel --- one for each possible kind of interaction
(activation, inhibition, no interaction).

\begin{table}[ht!]
\caption{Experiment 2 results. Average accuracy rates for estimation of the gene interaction parameter $a_{42}$.}
\centering
\begin{tabular}{ccccccc}
\midrule
$n$ & $p$ & $N$  & High Disp. & Low Disp.  \\
 &  & & $\phi=1$ & $\phi=5$ \\[2pt]
\midrule
\multirow{7}{*}{30} & \multirow{3}{*}{0.01} &  200  & 0.45 & 0.62  \\[2pt]
 &  & 1000& 0.58 & 0.71\\[2pt]
  &  & 5000 & 0.59 & 0.73\\[2pt]
\cmidrule{2-5}
 & \multirow{3}{*}{0.05} &  200 & 0.52 & 0.71\\[2pt]
 &  &  1000& 0.64 & 0.74 \\[2pt]
  &  &  5000& 0.67 & 0.75\\[2pt]
\midrule
\multirow{7}{*}{60} & \multirow{3}{*}{0.01} &  200  & 0.82 & 0.86\\[2pt]
 &  &  1000 & 0.87 & 0.93 \\[2pt] 
 &  &  5000 & 0.89 & 0.93\\[2pt]
\cmidrule{2-5}
 & \multirow{3}{*}{0.05} &  200  & 0.86 & 0.91\\[2pt]
 &  &  1000 & 0.89 & 0.95\\[2pt]
 &  &  5000 & 0.92 & 0.96 \\[2pt]
\midrule
\end{tabular}
\label{table:mixed}
\end{table}

Table~\ref{table:mixed} displays the average accuracy rate in the
estimation of the interaction type between \textit{Rb} and \textit{E2F}
over 100 different runs. We can observe that performance increases
with longer time series and larger number of particles, as
expected. The performance is better for larger transition noise.  The
reason is that large transition noise gets the system out of its
attractors more often and, as a result, helps the estimation
process. 

\begin{figure*}[ht!]
\centering
\includegraphics[width=150mm]{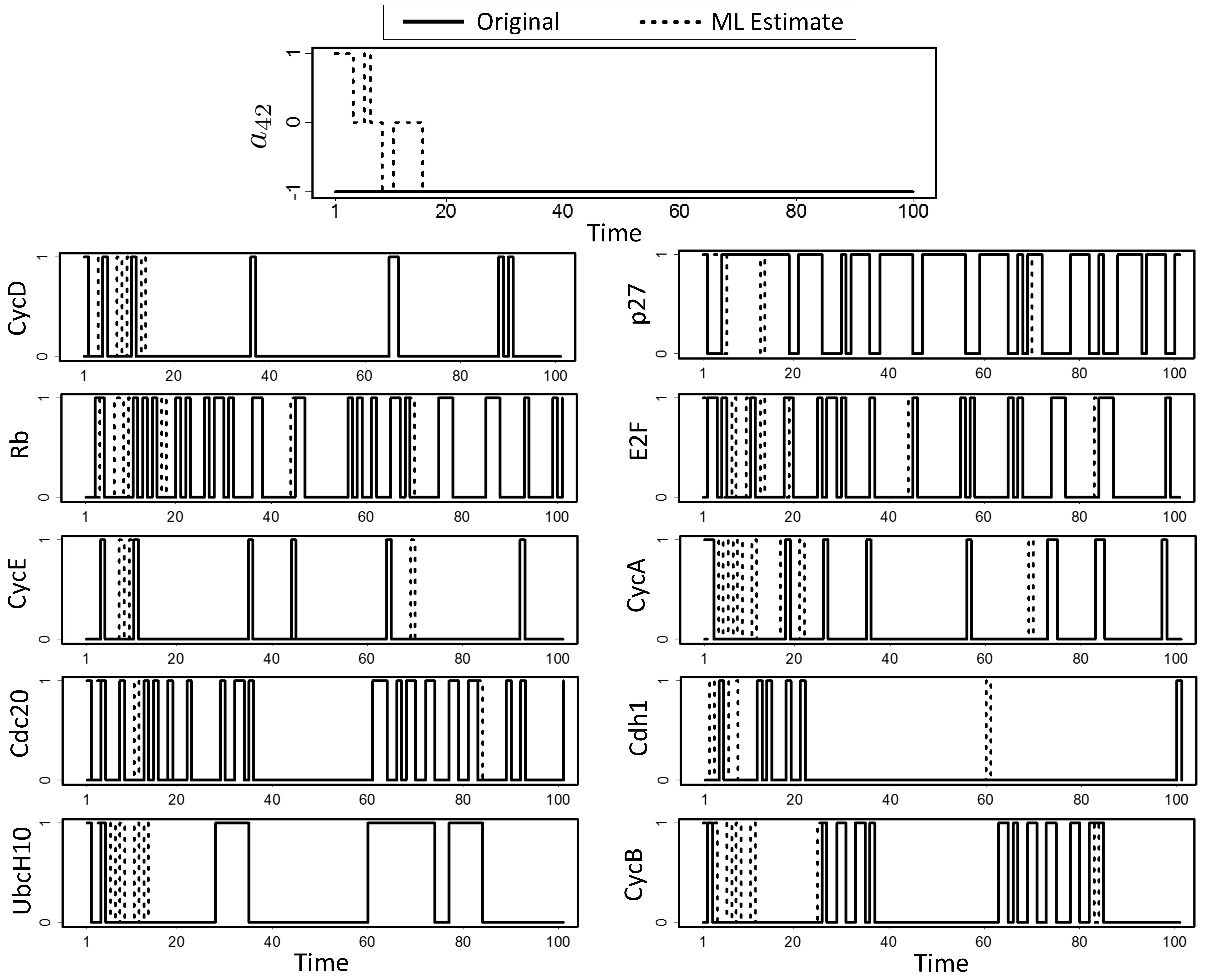}
\caption{Experiment 2: Estimated interaction type from \textit{Rb} to \textit{E2F} and true gene trajectories for
  single sample run of the experiment.}
 \label{fig:transition}
\end{figure*}

The evolution of estimated state and discrete-parameter for a single
sample run of the experiment, for transition noise $p=0.05$, $\phi=5$,
and $N=5000$ is displayed in Figure~\ref{fig:transition}. We can
observe that the discrete parameter $a_{42}$ is estimated correctly
after less than $20$ time step. In addition, we can see that the state
estimator of each gene eventually converges to the true state value.

\subsection{Experiment 3: Unknown Noise and Expression Parameters}

In the final experiment, the Boolean network topology (gene
interaction parameters $a_{ij}$ and biases $b_i$) is assumed to be
completely known, whereas the transition noise parameter $p$, the
baseline expression $\mu$, and differential expression $\delta_i$,
$i=1,\ldots,10$, are unknown. The inverse dispersion parameters are assumed to be $\phi_i = 5$, $i=1,\ldots,10$.

In order to assess continuous-parameter estimation accuracy, we define
the relative distance between estimated and true parameter values as
\beq\label{eq:NE1} \text{Relative
  Distance$(\hat{\theta})$}=\frac{\vert\hat{\theta}
  -\theta^*\vert}{R(\theta)}\,, \eeq where $\theta^*$ is the true
parameter value, and $R(\theta)$ is the range of parameter $\theta$
assumed in the M-step of the APF-CPMLA-BKS algorithm. Here, the range
is $R(p) = [0, 0.5]$ for the transition noise $p$, $R(\mu) = [0, 2]$
for the baseline expression $\mu$, and $R(\delta_i) = [0.1,10]$ for
the differential expression $\delta_i$, $i=1,\ldots,10$.

A new version of the ``augmented Lagrange method"
\cite{birgin2008improving} is used for optimization in the M-Step of
the particle-based EM algorithm. The gradient vector at each step is
computed based on Table~\ref{table:grad}. The procedure terminates
when the maximum of the absolute values of the changes in the
parameter estimates in two consecutive iterations gets smaller
than~$10^{-4}$.

\begin{figure*}[ht!]
\centering
\includegraphics[width=170mm]{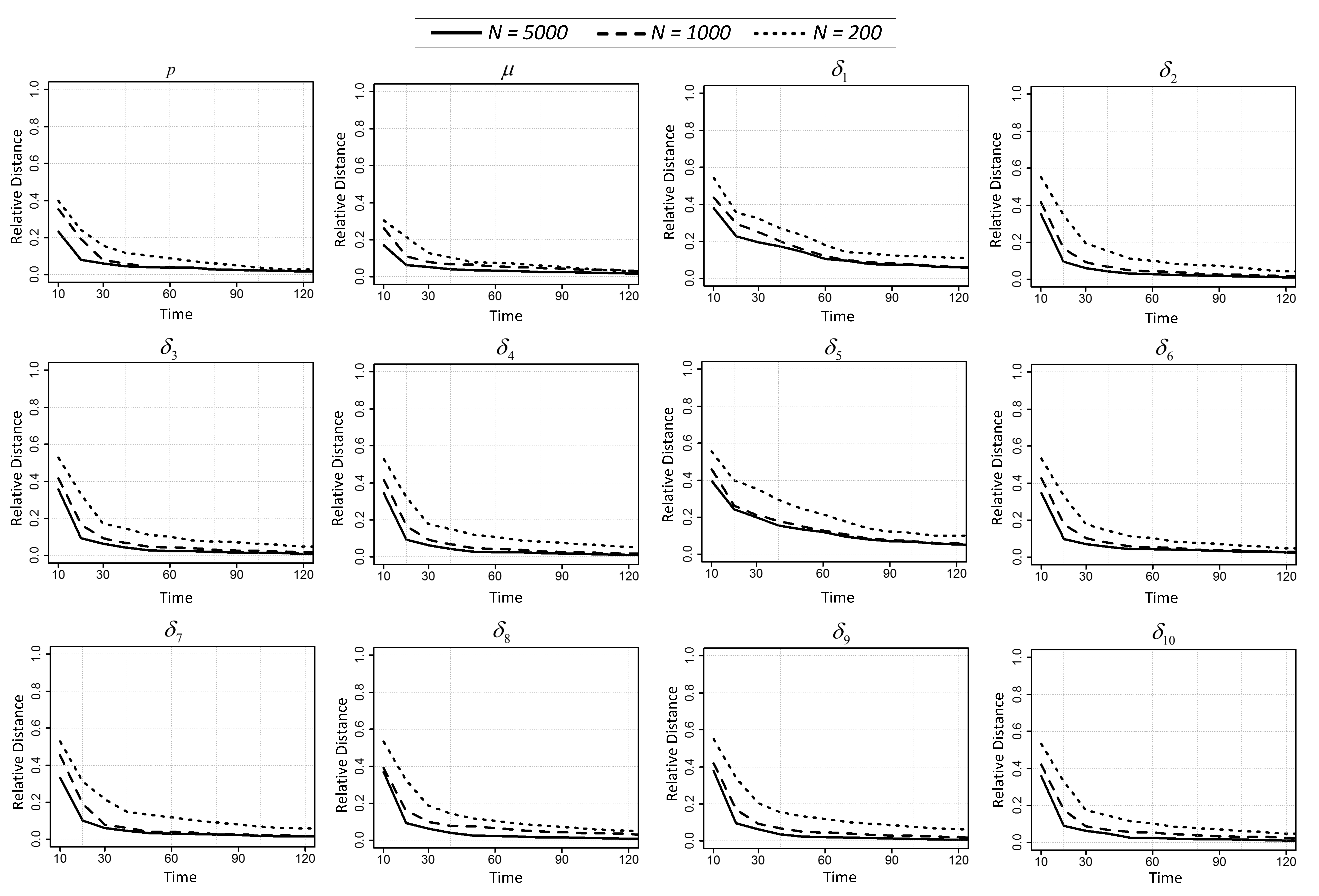}
\caption{Experiment 3: Average relative distance between
  estimated and true parameter values as a function of time series length.}
 \label{fig:perf}
\end{figure*}

The average relative distance between estimated and true parameter
values over 100 independent runs for different inverse dispersion
parameters and time series lengths are plotted in
Figure~\ref{fig:perf}. As expected, the performance of APF-CPMLA-BKS
improves steadily as time goes on. Performance improves by increasing
the number of particles; however, the curves get close to each other
as the length of the time series increases.  All curves show a
decreasing trend, which indicates that the parameter estimates become
arbitrarily close to the true values for sufficiently long time.

\section{Conclusion}
\label{sec:conc}

In this paper, we introduced approximate particle-based algorithms for
state and simultaneous state and parameter estimation for large
partially-observed Boolean dynamical systems. For approximate state
estimation, filtering and smoothing methods based on auxiliary
particle filtering (APF) were developed to approximate the optimal BKF
and BKS. These algorithms are called APF-BKF and APF-BKS, and are
original contributions of this work.

Moreover, we considered the case where some of the parameters may not
be known. In the discrete parameter case, an adaptive filter scheme is
developed based on APF-BKF algorithms running in parallel. For
continuous parameter problems, a particle-based EM algorithm for POBDS
was presented.

The methodology was applied to a model of Boolean gene regulatory
networks observed through RNA sequencing data. The numerical
experiments with a cell-cycle Boolean network demonstrated the ability
of the proposed methodologies to efficiently estimate the state and
also the parameters of the large Boolean regulatory network observed
through noisy measurements.

\section*{Acknowledgment}

The authors acknowledge the support of the National Science Foundation,
through NSF award CCF-1320884.

\bibliographystyle{ieeetr}
\bibliography{bibPF}

\end{document}